\documentclass[11pt,a4paper,reqno]{amsart}
\usepackage{amsthm,amsmath,amsfonts,amssymb,amsxtra,appendix,bookmark,euscript,delarray,dsfont,mathrsfs}
\usepackage{enumerate}
\usepackage[utf8]{inputenc}
\usepackage{xcolor}
\usepackage{graphicx}
\usepackage{upgreek}

\usepackage{tikz}
\usetikzlibrary{decorations.pathreplacing, calc}

\theoremstyle{plain}
\newtheorem{theorem}{Theorem}
\newtheorem{lemma}[theorem]{Lemma}
\newtheorem*{lemma*}{Lemma}

\newtheorem*{conjecture*}{Conjecture}

\theoremstyle{definition}
\newtheorem{definition}[theorem]{Definition}

\theoremstyle{remark}
\newtheorem{remark}[theorem]{Remark}

\DeclareMathOperator{\Tr}{Tr}

\def\bq{\begin{eqnarray}}
	\def\eq{\end{eqnarray}}
\def\bqq{\begin{eqnarray*}}
	\def\eqq{\end{eqnarray*}}

\def\epsilon{\varepsilon}

\newcommand{\abs}[1]{\left\lvert#1\right\rvert}

\renewcommand{\phi}{\varphi}
\newcommand\1{{\ensuremath {\mathds 1} }}

\def\N{\mathbb{N}}
\def\R{\mathbb{R}}

\def\Z{\mathbb{Z}}

\def\eps{\varepsilon}

\allowdisplaybreaks

\title[Lieb--Thirring inequalities with nearest-neighbor type interactions]{Lieb--Thirring inequalities for large quantum systems with inverse nearest-neighbor interactions}

\author[G. K. Duong]{Giao Ky Duong}
\address{Department of Mathematics, LMU Munich, Theresienstrasse 39, 80333 Munich, and Munich Center for Quantum Science and Technology (MCQST), Schellingstr. 4, 80799 Munich, Germany\\and 
	Institute of Applied Mathematics, University of Economics Ho Chi Minh City, Vietnam
}
\email{duong@math.lmu.de; kydg@ueh.edu.vn}

\author[P.T. Nam]{Phan Th\`anh Nam}
\address{Department of Mathematics, LMU Munich, Theresienstrasse 39, 80333 Munich, and Munich Center for Quantum Science and Technology (MCQST), Schellingstr. 4, 80799 Munich, Germany}
\email{nam@math.lmu.de}

\begin{document}
	
	\begin{abstract} We  prove an analogue of the Lieb--Thirring inequality for many-body quantum systems with the kinetic operator $\sum_i (-\Delta_i)^s$ and the  interaction potential of the form $\sum_i \delta_i^{-2s}$ where $\delta_i$ is the nearest-neighbor distance to the point $x_i$. Our result extends the standard Lieb--Thirring inequality for fermions and applies to quantum systems without the anti-symmetry assumption on the wave functions. Additionally,  we derive similar results for the Hardy--Lieb--Thirring inequality and obtain the asymptotic behavior of the optimal constants in the strong coupling limit.  
	\end{abstract}
	
	\maketitle

	%\tableofcontents
	
	\section{Introduction}
	
	The celebrated Lieb--Thirring inequality \cite{LiTh-75,LiTh-76} asserts that for any {\em anti-symmetric} normalized wave function $\Psi \in L^2(\R^{dN})$, we have
	\begin{equation}\label{eq:LT-fermions}
		\left\langle \Psi, \sum_{i=1}^N (-\Delta_{x_i})  \Psi \right\rangle \geq C_{\rm LT}(d) \int_{\R^d} \rho_\Psi^{1+\frac{2}{d}} (x) \,dx
	\end{equation}
	where $C_{\rm LT}(d)>0$ is a universal constant depending only on the dimension $d\ge 1$. Here the one-body density $\rho_\Psi \in L^1(\R^d)$ is defined by  
	\begin{equation}\label{defi-one-body-density}
		\rho_\Psi(x) := \sum_{j=1}^N \int_{\R^{(d-1)N}} \abs{\Psi(x_1,...,x_{j-1},x,x_{j+1},...,x_N)}^2 \prod_{i: i \neq j} \,dx_i. 
	\end{equation}
	In particular, $\rho_\Psi\ge 0$ and  $\int_{\R^d} \rho_\Psi(x) d x =N$, the total number of particles. The invention of the Lieb--Thirring inequality \eqref{eq:LT-fermions} was originally motivated by questions concerning the stability of matter \cite{LiTh-75}. In its dual form, it has also found significant applications in the spectral theory of the Schrödinger operator $-\Delta + V(x)$ on $L^2(\mathbb{R}^d)$. For detailed textbook introductions to these topics, we refer the reader to \cite{LiSe-10, FLW-23}. 
	
	\smallskip
	
	It is important to emphasize that the Lieb--Thirring constant $C_{\rm LT}(d)$ in \eqref{eq:LT-fermions}  is independent of $N$, making the bound very helpful for large quantum systems. In particular, the anti-symmetry assumption, which requires 
	\begin{equation} \label{eq:antisym}
		\Psi(x_1,...,x_i,...,x_j,...,x_N)=-\Psi(x_1,...,x_j,...,x_i,...,x_N),\quad \forall i\neq j, \quad \forall x_i, x_j \in \R^d,
	\end{equation}
	is a property reflecting the exclusion principle for \emph{fermionic particles}. Without the anti-symmetry condition \eqref{eq:antisym}, we  only have 
	\begin{equation}\label{eq-LT-bos}
		\left\langle \Psi, \sum_{i=1}^N (-\Delta_{x_i})   \Psi \right\rangle \geq \frac{C_{\rm GN}}{N^{2/d}} \int_{\R^d} \rho_\Psi^{1+\frac{2}{d}} (x) \,dx
	\end{equation}
	where $C_{\rm GN}$ is the sharp constant in the Gagliardo--Nirenberg interpolation inequality
	\begin{equation}\label{eq-GN}
		\Big( \int_{\R^d} |\nabla u|^2 \Big) \Big( \int_{\R^d} |u|^2 \Big)^{2/d} \ge  C_{\rm GN}(d) \int_{\R^d} |u|^{2(1+2/d)} , \quad \forall u\in H^1(\R^d). 
	\end{equation}
	The optimality of the constant $C_{\mathrm{GN}}N^{-2/d}$ in \eqref{eq-LT-bos} can be observed by choosing $\Psi = u^{\otimes N}$, a typical state of \emph{bosonic particles} (see, e.g., \cite[Eq. (6)]{LuNaPo-16} for an explanation).

		\smallskip
		
	Clearly, the general estimate \eqref{eq-LT-bos} is much weaker than the fermionic estimate \eqref{eq:LT-fermions} as $N \to \infty$. From a technical perspective, the loss of sharpness in the kinetic energy estimate \eqref{eq-LT-bos} is essentially comparable to the weakening of Sobolev's inequality in higher dimensions. The essence of \eqref{eq:LT-fermions} lies in the fact that, while Sobolev's inequality provides a quantitative justification of the uncertainty principle, it can be improved in higher dimensions under a suitable form of the exclusion principle.

	\smallskip

	For interacting quantum systems without the fermionic assumption, there are possibilities to obtain the Lieb--Thirring inequality provided that the repulsive interactions between particles  are sufficiently strong. This research direction was first proposed by Lundholm, Portmann and Solovej \cite{LuPoSo-15} where they showed that for any constant $\lambda>0$ and  any normalized wave function $\Psi$ in $L^2(\R^{dN})$, we have 
	\begin{equation}\label{eq-LT-potential-1}
		\left\langle \Psi, \left(\sum_{i=1}^N (-\Delta_{x_i}) + \sum_{1 \leq i<j \leq N} \frac{\lambda}{\abs{x_i-x_j}^2} \right)  \Psi \right\rangle 
		\geq \widetilde{K}_{\rm LT}(d,\lambda) \int_{\R^d} \rho_\Psi^{1+\frac{2}{d}} \,dx
	\end{equation}
with a constant $\widetilde{K}_{\rm LT}(d,\lambda)>0$ depending only on $d \ge 1$ and $\lambda>0$. This result  has direct applications to energy estimates for bosonic systems \cite{LuPoSo-15}. 

	\smallskip
	
The bound \eqref{eq-LT-potential-1} implies that the inverse-square interaction potential is strong enough to replace the anti-symmetry condition \eqref{eq:antisym} in playing the role of the exclusion principle for quantum particles. Later, it was proved in \cite{KogNam-21} that the optimal constant $\widetilde{K}_{\rm LT}(d,\lambda)$ in \eqref{eq-LT-potential-1} converges to the Gagliardo--Nirenberg constant ${C}_{\rm GN}(d)$ in Eq. \eqref{eq-GN} in the strong coupling limit, namely
	\begin{equation} \label{eq-LT-potential-2}
		\lim_{\lambda\to \infty} \widetilde{K}_{\rm LT}(d,\lambda) = C_{\rm GN}(d).
	\end{equation}
	Heuristically, \eqref{eq-LT-potential-2} captures the fact that if the interaction is too strong, every particle is forced to stay away from the others, and hence the many-body energy boils down to the one-body energy. 
	
	Note that  \eqref{eq-LT-potential-1} and \eqref{eq-LT-potential-2} were also extended to the fractional Laplacian $(-\Delta)^s$ with $s\in (0,\infty)$ and the corresponding two-body potential $|x-y|^{-2s}$. To be precise, it was proved in  \cite{LuNaPo-16} that 
	 for any constant $\lambda>0$ and  any normalized wave function $\Psi$ in $L^2(\R^{dN})$, we have 
	\begin{equation}\label{eq-LT-potential-1s}
		\left\langle \Psi, \left(\sum_{i=1}^N (-\Delta_{x_i})^s + \sum_{1 \leq i<j \leq N} \frac{\lambda}{\abs{x_i-x_j}^{2s}} \right)  \Psi \right\rangle 
		\geq \widetilde{K}_{\rm LT}(d,s,\lambda) \int_{\R^d} \rho_\Psi^{1+\frac{2s}{d}} \,dx
	\end{equation}
with a constant $\widetilde{K}_{\rm LT}(d,s,\lambda)>0$ depending only on $d \ge 1$, $s>0$ and $\lambda>0$. Moreover,  it was proved in \cite{KogNam-21} that the optimal constant $\widetilde{K}_{\rm LT}(d,s,\lambda)$ in \eqref{eq-LT-potential-1s} satisfies 
	\begin{equation} \label{eq-LT-potential-2s}
		\lim_{\lambda\to \infty} \widetilde{K}_{\rm LT}(d,s,\lambda) = C_{\rm GN}(d,s)
	\end{equation}
with the (fractional) Gagliardo--Nirenberg constant 
		\begin{equation} \label{eq:GN-constant-s}
			C_{\rm GN}(d,s)= \inf_{\substack{ u\in H^s(\R^d)\\[0.02cm] \|u\|_{L^2}=1}} \dfrac{\langle u, (-\Delta )^s u\rangle}{\int_{\R^d} |u|^{2(1+{\frac {2s} d})}}.
		\end{equation}
		In the case $s=1/2$, the estimates \eqref{eq-LT-potential-1s} and \eqref{eq-LT-potential-2s} are particularly relevant to the analysis of pseudo-relativistic particles. 
		
		\smallskip	
	
	The aim of the present paper is to extend the results in \eqref{eq-LT-potential-1s} and \eqref{eq-LT-potential-2s} to the case when the two-body interaction
	\begin{align} \label{eq:W-2}
		\sum_{1 \leq i<j \leq N} \frac{1}{\abs{x_i-x_j}^{2s}} 
	\end{align}
	is replaced by  the ($2s$-power) inverse nearest-neighbor interaction
	\begin{align}\label{eq:W-nearest}
		\sum_{i=1}^N  \frac{1}{\delta_i^{2s} (x_1,...,x_N)} ,\quad \delta_i(x_1,...,x_N) = \min_{j: j\ne i} |x_j-x_i|. 
	\end{align}

			\smallskip	
					
	The following is our first main result. 
	
	\begin{theorem}\label{thm:1-fractional}  Let  $d\in \mathbb{N}$, $s >0$ and  $\lambda>0$. Then there exists a constant $K_{\rm LT}(d, s, \lambda)>0$ such that for every $N \ge 1$  and every normalized wave function $\Psi\in L^2(\R^{dN})$ we have
		\begin{equation} \label{eq:LT-nearest-potential-1-fractional}
			\left\langle \Psi, \left(\sum_{i=1}^N (-\Delta_{x_i})^s   +  \sum_{i=1}^N  \frac{\lambda}{\delta_i^{2s} (x_1,...,x_N)}   \right)  \Psi \right\rangle 
			\geq K_{\rm LT}( d, s,  \lambda) \int_{\R^d} \rho_\Psi^{1+\frac{2s}{d}}.
		\end{equation}
		Moreover, there exist universal constants $C(d,s)>0,\lambda(d,s)$ and $k_1(d,s)>0$ such that the optimal constant in \eqref{eq:LT-nearest-potential-1-fractional} satisfies  
		\begin{equation} \label{eq:LT-nearest-potential-2-fr}
			C_{\rm GN}(d,s) \ge K_{\rm LT}(d,s, \lambda) \ge 
			\begin{cases} 
			\frac{\lambda}{C(d,s)} \quad &\text { if }\lambda\le \lambda(d,s),\\
			C_{\rm GN}(d,s)-\frac{C(d,s)}{\lambda^{k_1(s)}} >0, \quad  &\text { if }\lambda> \lambda(d,s).
			\end{cases}
		\end{equation}
		Here  $C_{\rm GN}(d,s)$ is the Gagliardo--Nirenberg constant defined in \eqref{eq:GN-constant-s}. If $s=1$, we can take $k_1 = \min\{\frac 1 3,  \frac 2 {3d}\}$. 
		\end{theorem}

Since the inverse nearest-neighbor interaction in \eqref{eq:W-nearest} is significantly smaller than the two-body interaction in \eqref{eq:W-2}, Theorem \ref{thm:1-fractional} provides stronger versions of \eqref{eq-LT-potential-1s} and \eqref{eq-LT-potential-2s}, thereby extending the previous results in \cite{LuPoSo-15, LuNaPo-16, KogNam-21}. Additionally, our result shows a quantitative convergence to the Gagliardo--Nirenberg constant $C_{\rm GN}(d, s)$ as $\lambda \to \infty$, which was not addressed in \cite{KogNam-21}. 

Historically, the inverse nearest-neighbor potential in \eqref{eq:W-nearest} arises in the study of the stability of matter, when one attempts to reduce the analysis of many-body quantum systems to one-body estimates; see \cite{FefdeL-86,LieYau-88} for pioneering works on relativistic systems and  \cite[Chapter 7]{LiSe-10} for the stability of non-relativistic matter. As we will explain below, using the inverse nearest-neighbor interaction in \eqref{eq:W-nearest} to gain a weak form of the exclusion principle is more physically relevant than using the two-body interaction in \eqref{eq:W-2}. %The results in Theorem \ref{thm:1-fractional} also have implications beyond those provided by previous works, as we explain below.

		\begin{remark}[Implication to the fermionic Lieb--Thirring inequality]	If $0<s<d/2$, then for every normalized anti-symmetric wave function $\Psi\in L^2(\R^{dN})$ we have
		\begin{equation}\label{eq:theo-3}
			\left\langle  \Psi, \left(\sum_{i=1}^N  (-\Delta_{x_i})^s \right) \Psi \right\rangle \ge  L(d,s) \left\langle \Psi, \sum_{i=1}^N \frac{1}{\delta_i^{2s}(x_1,...,x_N)} \Psi \right\rangle
		\end{equation}
		for a constant $L(d,s)>0$ depending only on $d$ and $s$. The bound \eqref{eq:theo-3} is of independent interest, and it goes back to  Fefferman--de la Llave \cite{FefdeL-86} and Lieb--Yau \cite{LieYau-88} for relativistic systems with $s=1/2$ in three dimensions; see Appendix \ref{app} for an extension to {$0<s<d/2$}. In this parameter regime, the new estimate \eqref{eq:LT-nearest-potential-1-fractional} in Theorem \ref{thm:1-fractional} implies the standard Lieb--Thirring inequality for fermions, while conceptually the corresponding estimate with the two-body interaction \eqref{eq:W-2} in \cite{LuPoSo-15, LuNaPo-16, KogNam-21} does not. 
			\end{remark}
			
\begin{remark}[Ground state energy in the thermodynamic limit] For $\lambda >0$ fixed, consider the Hamiltonian 
\begin{align}\label{eq:H-lambda-N-the}
 H_{\lambda, N} =  \sum_{i=1}^N (-\Delta_{x_i}) + \lambda \sum_{i=1}^N \delta_i^{-2}(x_1,...,x_N),
 \end{align}
acting  on $L^2([-L/2,L/2]^{dN})$ with Dirichlet boundary conditions. Theorem \ref{thm:1-fractional} implies that the ground state energy per unit volume $E_\lambda(\rho)$ of $H_{\lambda, N}$ in the thermodynamic limit 
$$N\to \infty, \quad L\to \infty, \quad NL^{-d} \to \rho>0,$$
 is bounded from below as 
\begin{align} \label{eq:optimal-hs-lower}
 E_\lambda(\rho) \ge C_{d,\lambda} \rho^{1+2/d}
 \end{align}
for a constant $C_{d,\lambda}>0$ depending only on $d$ and $\lambda$. It is not difficult to see that the energy asymptotic behavior of order $\rho^{1+2/d}$ is optimal in the low density limit $0< \rho \ll 1$. For example, in the non-relativistic case $s=1$ in three dimensions, by using the ground state $\Psi_{\rm hs}$ of the Bose gas with the hard-sphere interaction  
$$
V(x-y)=  \begin{cases} 0\quad &\text{ if }  |x-y| > a,\\
+\infty \quad & \text{ if }  |x-y|\le a
\end{cases}
$$
as a trial state, we find that  
\begin{align} \label{eq:optimal-hs}
E_\lambda(\rho) \le 4\pi a \rho^2 (1 + o(1)_{a\rho^3\to 0}) + \lambda \rho a^{-2} 
\end{align}
where the first term comes from Dyson's upper bound \cite{Dyson-57}, and the second term arises from the fact that $\langle \Psi_{\rm hs}, \delta_i^{-2}(x_1,...,x_N) \Psi_{\rm hs}\rangle \le a^{-2}$. Optimizing over $a>0$ in \eqref{eq:optimal-hs} (taking $a\sim \rho^{-1/3}$), we obtain $E_\lambda(\rho)\le O(\rho^{5/3})$, which agrees with the lower bound in \eqref{eq:optimal-hs-lower} up to a constant factor. Determining $E_\lambda(\rho)$ exactly to the leading order is an interesting problem, which is left open. 

In contrast, if we replace the inverse nearest-neighbor interaction in \eqref{eq:H-lambda-N-the} by the two-body interaction $|x-y|^{-2}$, then the thermodynamic limit do not exists in dimensions $d\ge 2$ due to the long-range order of the interaction. If $d\ge 3$, it is straightforward to see that for every normalized wave function $\Psi\in L^2([-L/2,L/2]^{dN})$ we have
$$ \frac{1}{L^{d}}   \left\langle  \Psi, \sum_{i < j} \frac{1}{|x_i-x_j|^2}   \Psi  \right\rangle \ge  \frac{1}{L^{3}}   \left\langle  \Psi, \sum_{i < j} \frac{1}{L^2}   \Psi  \right\rangle = \frac{N(N-1)}{2L^{d+2}} \to \infty$$
when $N= L^d \rho\to \infty$.   
\end{remark}

%\smallskip 
%\noindent 
%Now, we define the ground state energy of Hamiltonian $H_{\lambda, N}$  
%\begin{equation}
%	E_D(N,L,\lambda) = \inf {\rm spec } H_{\lambda, N}  = \inf_{ \substack{\Psi \in H^1(\Lambda_L),\\\| \Psi\|_{L^2(\Lambda)} =1}    }   \langle \Psi,H_N \Psi \rangle.
%\end{equation}
%\noindent 	
%
%		
%
%
%
%	
%		Since the inverse nearest-neighbor interaction is much smaller than the full two-body interaction \eqref{eq:W-2}, Theorem \eqref{thm:1-fractional} extends significantly the results in \cite{LuPoSo-15,LuNaPo-16,KogNam-21}. In addition, our result  shows an explicit behavior of the optimal constant in the regime of large $\lambda$.      Moreover, while conceptually the estimates with the full two-body interaction \eqref{eq:W-2} are not enough to imply the standard Lieb--Thirring inequality for fermions, the new estimate \eqref{eq:LT-nearest-potential-1-fractional} does. To be precise, if $0<s<d/2$, then for anti-symmetric wave functions we have
%%		\begin{equation}\label{eq:theo-3}
%%			\left\langle  \Psi, \left(\sum_{i=1}^N  (-\Delta_{x_i})^s \right) \Psi \right\rangle \ge  L(d,s) \left\langle \Psi, \sum_{i=1}^N \frac{1}{\delta_i^{2s}(x_1,...,x_N)} \Psi \right\rangle,  
%%		\end{equation}
%		for a constant $L(d,s)>0$. The bound \eqref{eq:theo-3} is of independent interest; it goes back to Fefferman and de la Llave \cite{FefdeL-86} as well as Lieb and Yau \cite{LieYau-88} for the relativistic kinetic operator $s=1/2$ in three dimensions, and an extension to {$0<s<d/2$} will be given in Appendix \ref{app}.  
%	\end{remark}
%	
	
	\smallskip 
	
	In the next result, we prove a Hardy--Lieb--Thirring inequality with  the inverse nearest-neighbor interaction. Here we take $0<s<d/2$ and replace the kinetic operator $(-\Delta)^s$ by the Hardy--Schr\"odinger operator  
	$$
	(-\Delta)^s - \frac{\mathcal{C}_{d,s}}{|x|^{2s}} \ge 0 \quad \text{ on }L^2(\R^d),
	$$
with the optimal constant \cite[Eq. (2.6)]{Herbst-CMP-1997}
	\begin{equation}\label{Frac-Hardy-constant}
		\mathcal{C}_{d,s} :=2^{2 s}\left(\frac{\Gamma((d+2 s) / 4)}{\Gamma((d-2 s) / 4)}\right)^2. 
	\end{equation}
	
	\begin{theorem}\label{thm:2-fractional}    Let $d\in \mathbb{N}$, $0 < s <d/2$ and  $\lambda>0$.  Then there exists a constant $K_{\rm HLT}(d, s, \lambda)>0$ such that for every $N\ge 1$ and every normalized wave function $\Psi \in L^2(\R^{dN})$, we have 
		\begin{equation} \label{eq:HLT-nearest-potential-1-fractional}
			\left\langle \Psi, \left(\sum_{i=1}^N \left( (-\Delta_{x_i})^s - \frac{\mathcal{C}_{d,s}}{|x_i|^{2s}}  \right)+ \sum_{i=1}^N  \frac{\lambda}{\delta_i^{2s}(x_1,...,x_N)}    \right)  \Psi \right\rangle 
			\geq K_{\rm HLT}(  d, s,  \lambda) \int_{\R^d} \rho_\Psi^{1+\frac{2s}{d}}. 
		\end{equation}
		Moreover, there exist universal constants $C(d,s)>0,\lambda(d,s)$ and $k_2(d,s)>0$ such that the optimal constant in \eqref{eq:HLT-nearest-potential-1-fractional} satisfies  
		\begin{equation} \label{eq:LT-nearest-potential-2-fr-hardy}
			C_{\rm HGN}(d,s) \ge K_{\rm HLT}(d,s, \lambda) \ge 
			\begin{cases} 
			\frac{\lambda}{C(d,s)} \quad &\text { if }\lambda\le \lambda(d,s),\\
			C_{\rm HGN}(d,s)-\frac{C(d,s)}{\lambda^{k_2(s)}}>0, \quad  &\text { if }\lambda> \lambda(d,s).
			\end{cases}
		\end{equation}
		Here  $C_{\rm HGN}(d,s)$ is the Hardy--Gagliardo--Nirenberg constant 
		\begin{equation} \label{constant-HGN}
			C_{\rm HGN}(d, s):= \inf_{\substack{u\in H^s(\R^d)\\ \|u\|_{L^2}=1}
			} \dfrac{\left\langle u, ((-\Delta)^s -\frac{\mathcal{C}_{d,s}}{|x|^{2s}}) u\right\rangle}{\int_{\R^d} |u|^{2(1+2s/d)}}.
		\end{equation}
		If $s=1$, we can take $k_2= 1/(2d)$. 
		\end{theorem}

	The results in Theorem \ref{thm:2-fractional} can be formally interpreted as the bosonic analogue of the Hardy–Lieb–Thirring inequality for fermions  established in \cite{EkhFra-06,FraLieSei-08,Frank-09}. Note that \eqref{eq:HLT-nearest-potential-1-fractional} is stronger than \eqref{eq:LT-nearest-potential-1-fractional}, although it is unclear whether \eqref{eq:HLT-nearest-potential-1-fractional} implies the fermionic Hardy–Lieb–Thirring inequality  since the extension of \eqref{eq:theo-3} to the Hardy--Schrödinger operator is not available. 
	
	Moreover, the asymptotic formula \eqref{eq:LT-nearest-potential-2-fr-hardy} in Theorem \ref{thm:2-fractional} improves the corresponding result in \cite{KogNam-21}, which was obtained for the two-body interaction $|x-y|^{-2s}$ without providing an explicit convergence rate.  The power $k_2(s)$ of $\lambda$ in  \eqref{eq:LT-nearest-potential-2-fr-hardy}, as well as the power $k_1(s)$ in  \eqref{eq:LT-nearest-potential-2-fr},  is not necessarily optimal, and determining the sharp power remains an interesting issue.

\smallskip
\noindent
\textbf{Proof strategy:} Our general strategy to  prove Theorem \ref{thm:1-fractional} and Theorem \ref{thm:2-fractional} goes back to the microlocal techniques introduced by Lundholm and Solovej \cite{LuSo-13,LuSo-13b, LuSo-14}, which were inspired by the Dyson--Lenard proof of the stability of matter \cite{	DyLe-67,DyLe-68} and  further developed in   \cite{FrSe-12,LuPoSo-15,LuNaPo-16,Nam-17,LaLu-18, LuSe-18,LaLuNa-19,KogNam-21,Nam-22}. 

In particular, we will adopt the approach in \cite{KogNam-21}, but the treatment of the inverse nearest-neighbor interaction requires subtle refinements. 

\begin{itemize}

\item First, we will revisit and improve the localization procedure in \cite{KogNam-21}. In particular, we will derive a quantitative local uncertainty principle by using the detailed geometric property of the localized domains. 

\item Second, since  the inverse nearest-neighbor interaction is much weaker than the two-body interaction $|x-y|^{-2s}$, it is more challenging to obtain a local  version of the exclusion principle. In this aspect, we will start by using an idea from \cite{KogNam-21} to  split the interaction energy into different length scales, but then we derive a new local exclusion principle by using the Besicovitch covering lemma. 

% we combine the local uncertainty and exclusion estimates using the Besicovitch covering lemma. This idea originates from the seminal work of Rozenblum \cite{R76} on the number of negative eigenvalues of Schr\"odinger operators and was recently used in \cite{Nam-22} to provide a new, direct proof of the standard (fermionic) Lieb–Thirring inequality. However, in these works, the Besicovitch covering lemma often results in a significant loss of the constant factor. In contrast, we employ this tool to obtain \eqref{eq:LT-nearest-potential-2-fr} and \eqref{eq:LT-nearest-potential-2-fr-hardy}, with sharp constants. 
%
%
%In this aspect, we will employ an idea from \cite{LaLuNa-19} to establish a weak form of the exclusion principle. 
%
%
%
%\item 
%Third, unlike in \cite{KogNam-21}, we combine the local uncertainty and exclusion estimates using the Besicovitch covering lemma. This idea originates from the seminal work of Rozenblum \cite{R76} on the number of negative eigenvalues of Schr\"odinger operators and was recently used in \cite{Nam-22} to provide a new, direct proof of the standard (fermionic) Lieb–Thirring inequality. However, in these works, the Besicovitch covering lemma often results in a significant loss of the constant factor. In contrast, we employ this tool to obtain \eqref{eq:LT-nearest-potential-2-fr} and \eqref{eq:LT-nearest-potential-2-fr-hardy}, with sharp constants. 

\end{itemize}

These new ingredients help us to obtain good quantitative estimates, which are important for establishing the asymptotic estimates \eqref{eq:LT-nearest-potential-2-fr} and \eqref{eq:LT-nearest-potential-2-fr-hardy}, with sharp constants. We hope the new techniques presented here will be helpful in studying the standard Lieb--Thirring inequality in the cases where the conjectured optimal constant coincide with the Gagliardo–Nirenberg constant (e.g., the Lieb–Thirring conjecture in one and two dimensions).

\smallskip
\noindent
\textbf{Structure of the paper:}  In Section \ref{sec:Preliminaries} we collect some preliminary estimates concerning the (fractional) Sobolev spaces. In Section \ref{section-subcovering} we revisit the localization procedure in \cite{KogNam-21} and establish a new feature which is important to improve error estimates. Then we derive local versions of the uncertainty and exclusion principles in  Section \ref{section-local-uncertainty-principles} and Section \ref{section-exclusion-principle}. Finally, we prove Theorem \ref{thm:1-fractional} in Section \ref{section-proo-theorem-1} and prove  Theorem \ref{thm:2-fractional} in Section \ref{sec:HLT}. All of the estimates here hold for all wave functions without any symmetry assumption. Specific estimates for fermions, including \eqref{eq:theo-3},  are discussed separately in Appendix \ref{app}.  

% by assuming certain technical results in Section \ref{section-covering-Laplace}. Additionally, we provide the proof of Theorem \ref{thm:2-fractional} by assuming  technical results in Section \ref{section-proof-theorem-2}.  In Section \ref{section-prove-theorem-conver}, we will give the complete proof of Theorem \ref{theo:impro-conver}. Next, Section \ref{section-prove-theorem-dominate} is devoted to the proof of Theorem \ref{theo-3}.
%In Section \ref{section-local-uncertainty-principles}, it  is devoted to  establish local uncertainty principles. Furthermore, we  prove local exclusion principles in Section \ref{section-exclusion-principle}. Finally, in Section \eqref{app}, we  provide some necessary comparison between Sobolev norms.

%\iffalse 
%Note that \eqref{eq:HLT-nearest-potential-1} and \eqref{eq:HLT-nearest-potential-2} are improvements of \cite[Theorem 2]{LuNaPo-16} and \cite[Theorem 2]{KogNam-21}, respectively. 
%
%\bigskip
%\noindent
%{\bf Question 1:} Can we extend these results to the case of fractional Laplacian? {\bf It should be true, although the proof may be more technical to read!}   
%
%\bigskip
%\noindent
%{\bf Question 2:} Can we show the following analogue of \eqref{eq:FS-LY}: 
%\begin{equation} \label{eq-question}
%	\left\langle \Psi, \left(\sum_{i=1}^N \left( -\Delta_{x_i} - \frac{(d-2)^2}{4|x_i|^2} \right)\right)  \Psi \right\rangle 
%	\geq C \left\langle \Psi, \sum_{i=1}^N \frac{1}{\delta_i^2}  \Psi \right\rangle ?
%\end{equation}
%It would be very interesting. But {\bf \eqref{eq-question} may be wrong?} 
%
%
%\fi 

\subsection*{Acknowledgments} This project was initiated in part during the authors' visit to the Vietnam Institute for Advanced Study in Mathematics (VIASM) in August 2022. We extend our gratitude to the staffs of the VIASM for their warm hospitality. PTN would like to thank Robert Seiringer for helpful remarks on the inverse nearest-neighbor interaction, and thank Ngo Quoc Anh, Nguyen Van Hoang, and Trinh Viet Duoc for inspiring discussions during his time at the VIASM. Partial support from the Deutsche Forschungsgemeinschaft (DFG,
German Research Foundation) through Germany’s Excellence Strategy EXC - 2111 - 390814868 and
through TRR 352 – Project-ID 470903074 is acknowledged.

\section{Global and local Sobolev norms}\label{sec:Preliminaries}

In this preliminary section  we  recall the setting of the fractional Sobolev spaces in $\R^d$ and in bounded domains. In particular, we will collect some elementary but helpful estimates comparing Sobolev norms. 

%We will use the microlocal techniques first introduced by Lundholme and Solovej \cite{LuSo-13} (see also \cite{KogNam-21} and \cite{LuNaPo-16}) to prove local 

 For each $s > 0$, we write $s = m + \sigma $, where $m \in \{0,1,2,...\}$ and $\sigma=\sigma(s) \in [0,1)$. Thus $m=[s]$, the largest integer which is smaller than or equal to $s$. The operator $(-\Delta)^s$ on $L^2(\R^d)$ can be defined using the quadratic form expression 
\begin{equation}\label{iden-u-Delta-s-u}
	\langle u, (-\Delta)^s u \rangle = \sum_{|\alpha| =m}  \frac{m!}{\alpha!} \langle \partial^\alpha, (-\Delta)^\sigma \partial^\alpha \rangle.
\end{equation} 
Here we denote 
$$ \alpha! = \alpha_1! ... \alpha_n!  \text{ and }  \partial^\alpha = \partial_{x_1}^{\alpha_1}...\partial^{\alpha_d}_{x_d}.$$
for any multi-index $\alpha = (\alpha_1,...,\alpha_d)\in \{0,1,...\}^d$    and  $x = (x_1,...,x_d) \in \R^d$, and for $ \sigma \in (0,1)$ we have the following definition
\begin{equation}
	\langle u, (-\Delta )^\sigma u \rangle  = c_{d,\sigma} \int_{\R^d}\int_{\R^d} \frac{|u(x) - u(y)|^2}{|x-y|^{d+2\sigma}} dx dy  \quad \text{ with }  c_{d,\sigma} = \frac{2^{2\sigma -1} \Gamma(\frac{d}{2} + \sigma)}{\pi^\frac{d}{2} \left|\Gamma(-\sigma) \right|} 
\end{equation}
(we set $(-\Delta)^\sigma \equiv \1$, the identity, if $\sigma =0$). Thus,   \eqref{iden-u-Delta-s-u} can be rewritten as follows
\begin{equation}
	\left\langle u,(-\Delta)^s u\right\rangle=
	\begin{cases}
	\sum_{|\alpha|=m} \frac{m !}{\alpha !} \int_{\R^d} |\partial^\alpha u(x)|^2 d x\quad \text{ if }\sigma(s)=0,\\
	c_{d, \sigma} \sum_{|\alpha|=m} \frac{m !}{\alpha !} \int_{\mathbb{R}^d} \int_{\mathbb{R}^d} \frac{\left|\partial^\alpha u(x)-\partial^\alpha u(y)\right|^2}{|x-y|^{d+2 \sigma}} d x d y\quad \text{ if }0<\sigma(s)<1.
	\end{cases}
\end{equation}
Now, let $\Omega$ be a domain in $\R^d$. We define $(-\Delta)^s_{|\Omega}$  as an operator on $L^2(\R^d)$ by Friedrichs' method via the  quadratic form
\begin{equation}\label{defi-seminorm-dot-H-s}
	\langle u, (-\Delta)^s_{|\Omega} u \rangle  = \|u\|^2_{\dot{H}^s(\Omega )} = \begin{cases}
		\sum_{|\alpha|=m} \frac{m !}{\alpha !} \int_{\Omega}\left|\partial^\alpha u\right|^2 d x, \quad \text { if } \sigma(s) = 0 \\ 
		c_{d, \sigma} \sum_{|\alpha|=m} \frac{m !}{\alpha !} \int_{\Omega} \int_{\Omega} \frac{\left|\partial^\alpha u(x)-\partial^\alpha u(y)\right|^2}{|x-y|^{d+2 \sigma}} d x d y, \quad \text { if } 0<\sigma<1.  
		\end{cases}
\end{equation}
We also define 
\begin{equation}\label{defi-norm-H-s-Omega}
	\|u\|_{H^s(\Omega)}^2:=\|u\|_{\dot{H}^s(\Omega)}^2+\sum_{|\alpha| \leq m} \int_{\Omega}\left|\partial^\alpha u\right|^2 d x.
\end{equation}

\smallskip 
We define the fractional Sobolev space $H^s(\Omega)$  as the space of all $u \in L^2(\Omega)$ satisfying $\| u\|_{H^s(\Omega)} < \infty$. When the boundary of $\Omega$ is sufficiently regular, we can extend any function in $H^s(\Omega)$ to $H^s(\R^d)$. To be precise, we recall  the following definition from \cite[page 240]{leoni2023}. 
  
\begin{definition}\label{defi-class-C-m-1} Let $d,M,m\in \mathbb{N}$ and $\eta,L >0$. Let  $\Omega \subset \mathbb{R}^d$ be an open subset.  The boundary  $\partial \Omega$   is called   $(\eta, L, M)$-uniformly of class $C^{m,1}$ if there exists a locally finite,  countable open cover $\left\{\Omega_n\right\}$ of $\partial \Omega$ such that the following properties are satisfied:
	\smallskip
	\begin{itemize}
		\item[(i)] if $x \in \partial \Omega$, then $B(x, \eta) \subset \Omega_n$ for some $n \in \mathbb{N}$;\\[0.3cm]
		\item[(ii)] each point of $\mathbb{R}^N$ is not contained in more than $M$ of the $\Omega_n$;\\[0.3cm]
		\item[(iii)] for each $n$, there exists a rigid motion $T_n: \R^{d} \to \R^d$, and a $C^{m,1}$ function $f_n: \R^{d-1} \to \R$, with $ \|f\|_{C^{m,1}(\R^{d-1})} \leq L$ such that 
		$$
		\Omega_n \cap \Omega=\Omega_n \cap T_n^{-1}\left( \left\{\left(y^{\prime}, y_d\right) \in \mathbb{R}^{d-1} \times \mathbb{R}: y_d>f\left(y^{\prime}\right)\right\}\right) .
		$$
	\end{itemize}
\end{definition}

We have  the following well-known result (see, e.g., \cite[Theorem 13.17]{Leoni-book2017t} for $s\in \mathbb{N}$ and  \cite[Theorem 11.57]{leoni2023}  for $s\notin \mathbb{N}$).

\begin{lemma}[Extension of Sobolev spaces]\label{lemma-extension-operator}
	Let $M,d \in \mathbb{N}$ and $\eta,L>0$. Let $ s \ge 0$ and $m = [s]$. Let  $\Omega \subseteq \mathbb{R}^d$ be an open set whose boundary is  $(\eta, L, M)$-uniformly class of $C^{m,1}$.  Then,  there exists a continuous linear operator $T: H^{s}(\Omega) \rightarrow H^{s}\left(\mathbb{R}^d\right)$ such that for all
	$u \in H^{s}(\Omega)$ we have $ E(u)=u$  a.e. on $\Omega$, and
	$$
	\|T(u)\|^2_{H^s\left(\mathbb{R}^d\right)} \leq \textbf{C}^*\|u\|^2_{H^s(\Omega)},
	$$
	where $\textbf{C}^*:=C(d, s,\eta, L, M)$ depending only on $d,s,\eta,L,$ and $ M$.
\end{lemma}

\smallskip 
Consequently,   we obtain the  following estimates which are helpful for our localization procedure. 

\begin{lemma}[Comparison of Sobolev norms] \label{lemma-compa-Sobolev-Lipschitz} Let $ d,M\in \mathbb{N}$ and $\eta,L>0$. Let $s>t  >0$ and $m = [s]$.  Let $\Omega \subset \R^d$ be an open subset whose boundary is  $(\eta, L, M)$-uniformly class of $C^{m,1}$. Then, for all $u\in H^s(\Omega)$, we have
	\begin{equation}\label{compa-H-s-H-dot-s-improve}
		\| u\|^2_{H^{s}(\Omega)}  \le \mathcal{C}_1 \left(  \|u\|^2_{\dot{H}^s(\Omega)} + \| u\|^2_{L^2(\Omega)}        \right) .
	\end{equation}
	Moreover, for all  $\xi \in (0,1)$, 
	\begin{equation}\label{compa-H-t-dot-H-s-impro}
		\|u\|^{2}_{H^{t}(\Omega)} \le \xi \| u\|^2_{\dot{H}^s(\Omega)} + \frac{\mathcal C_1}{\xi^{\frac{t}{s-t}} }\|u\|^2_{L^2(\Omega)}.
	\end{equation}
	Here the constant $\mathcal C_1 = C(d, s, \eta, L, M)$ depends only on $d, s, \eta, L, $ and $M$.%compare-So
\end{lemma}
\begin{proof}
	The idea of the proof is similar to  \cite[Lemma 5]{KogNam-21}. First, let us consider   \eqref{compa-H-s-H-dot-s-improve}. Since the proof for the integer case is easier, we will only explain  the case  $\sigma =\sigma(s) \in (0,1)$. According to    \eqref{defi-norm-H-s-Omega}, it  is sufficient to prove 
	\begin{equation}\label{esti-norm-H-le-dor-s-L-2}
		\|u\|_{{H}^m(\Omega)}^2  \le  C  \left( \|u\|^2_{\dot H^s(\Omega)} +  \|u\|^2_{L^2(\Omega)} \right),
	\end{equation}
	where $C= C(d, s, \eta, L, M)$ and $u \in H^s(\Omega)$. By using    
	H\"older's  inequality in Fourier space,  one has
	\begin{eqnarray}
		\|f\|_{H^m(\R^d)}^2 \le C(d)\|f\|_{H^s(\R^d)}^\frac{2m}{s} \|f\|^{2\left( 1-\frac{m}{s} \right)}_{L^2(\R^d)}, \quad \forall f \in H^s(\R^d) \label{control-f-H-m-H-sL2}. 
	\end{eqnarray}
	In addition to that, we apply the  extension result in Lemma \ref{lemma-extension-operator} in combining with     Young's inequality and \eqref{control-f-H-m-H-sL2},   we obtain 
	\begin{align*}
		\|u\|^{2}_{H^m(\Omega)} &\le C(d)\|T u\|_{H^s(\R^d)}^\frac{2m}{s} \|Tu\|^{2\left( 1-\frac{m}{s} \right)}_{L^2(\R^d)} \\
		& \le C(d) \left( a \|Tu\|_{H^s(\R^d)}^2 \right)^{\frac{m}{s}} \left( a^{-\frac{m}{s-m}} \|Tu\|_{L^2(\R^d)}^2 \right)^{1-\frac{m}{s}} \\
		&\le C(d) a  \|Tu\|_{H^s(\R^d)}^2  + Ca^{-\frac{m}{s-m}} \|Tu\|^2_{L^2}(\R^d)\\
		&\le C(d, s, \eta, L, M) \left( a \| u \|_{H^s(\Omega)} + a^{-\frac{m}{s-m}} \|u\|^2_{L^2(\Omega)} \right) ,\\
		&\le  C(d, s, \eta, L, M) \left( a \| u \|_{\dot H^s(\Omega)} + a \|u\|_{H^m(\Omega)} + a^{-\frac{m}{s-m}} \|u\|^2_{L^2(\Omega)} \right)
	\end{align*}
	for  every $a >0$. This implies  
	\begin{align*}
		\left( 1 - Ca \right) \|u\|^2_{H^m(\Omega)} \le C a \|u\|^2_{\dot H^s(\Omega)} + C a^{-\frac{m}{s-m}} \|u\|^2_{L^2(\Omega)},
	\end{align*}
	where  $C = C(d, s, \eta, L, M)  $. 	Thus, by taking $a>0 $ small enough, we obtain \eqref{esti-norm-H-le-dor-s-L-2}, and \eqref{compa-H-s-H-dot-s-improve} follows. 
	
Next, we consider  \eqref{compa-H-t-dot-H-s-impro}. For $t \in (0,s)$, we have
	\begin{equation}
		\| f\|_{\dot H^t(\R^d)}^2 \le C(d,s,t) \|f\|^{2 \frac{t}{s}}_{\dot H^s(\R^d)} \| f\|^{2\left( 1 -\frac{t}{s} \right)}_{L^2(\R^d)}, \forall u \in \dot H^s(\R^d), t \in (0,s),\label{dot-H-t-dot-H-s}
	\end{equation}
	which  is  similar to  \eqref{control-f-H-m-H-sL2}. Then by the same technique as above, we have 
	\begin{equation*}
		\begin{aligned}\|u\|_{\dot{H}^t(\Omega)}^2 & \leq\|T u\|_{\dot{H}^t\left(\mathbb{R}^d\right)}^2 \le C(d,s,t)\|T u\|_{\dot{H}^s\left(\mathbb{R}^d\right)}^{2 \frac{t}{s}}\|T u\|_{L^2(\R^d)}^{2\left(1-\frac{t}{s}\right)}  \\
			& \leq C(d, s, \eta, L, M) \left(a\|u\|_{H^s(\Omega)}^2+a^{-\frac{t}{s-t}}\|u\|_{L^2(\Omega)}^2\right).\end{aligned}
	\end{equation*}
for all   $ a   >0  $. By   taking  $a   = \xi(1+C(d, s, \eta, L, M))^{-1}$ we obtain   \eqref{compa-H-t-dot-H-s-impro} follows. \end{proof}

\section{Construction of covering sub-cubes}\label{section-subcovering}

In this section, we revisit the localization procedure in \cite{KogNam-21} and introduce a new property that allows us to keep track several quantitative error estimates.

\medskip

Let $N \in \mathbb{N}$,   $\delta \in (0,1)$ and    
$\varepsilon \in  \{\frac{1}{2}, \frac{1}{3}\}$. For any given function $0 \le \rho \in L^1(\R^d)$ supported in  $ \left[-\frac{1}{2}, \frac{1}{2} \right]^d$, we construct  a covering sub-cubes for $\left[-\frac{1}{2}, \frac{1}{2} \right]^d$ by induction as follows.

\smallskip 
\noindent
{\bf Initial step.} When $n =0$, we  define 
$$ G^0 =  \left\{ \left[ -\frac{1}{2}, \frac{1}{2} \right]^d \right\} = G^{0,0} \cup G^{0,1} \cup G^{0,2} \text{ where }   G^{0,0} = G^{0,1} = \emptyset. $$  

\smallskip 
\noindent 
{\bf Induction step.} Let  $G^{n} = G^{n, 0} \cup G^{n, 1} \cup G^{n, 2}$ be the collection of sub-cubes of side length $\varepsilon^n$, obtained from the  $n^{\rm th}$ step. If $G^{n,2}$ is empty, then we stop the procedure. If $G^{n,2}$ is non-empty, then  we  divide  each  cube $Q \in G^{n,2}$  into  new sub-cubes  of side length $\varepsilon^{n+1}$. Now we construct the collections 
\begin{equation*}
	G^{n+1} = G^{n+1, 0} \cup G^{n+1, 1} \cup G^{n+1, 2}
\end{equation*}
of these sub-cubes of side length $\varepsilon^{n+1}$ as follows: 
\begin{itemize}
	\item  $G^{n+1,0}$ is  the collection of  all sub-cubes  in $G^{n+1}$ satisfying
	\begin{equation}\label{condi-Q-in-Q-n-0}
		\int_{Q} \rho \le \delta. 
	\end{equation}
	\item 	For $G^{n+1,1}$, we decompose the sub-cubes in $G^{n+1} \backslash  G^{n+1,0}$ into \textbf{clusters} $K$, which are connected components $K$ in graphical sense, where the two sub-cubes $Q_1$ and $Q_2$ are connected if ${\rm dist}(Q_1,Q_2)=0$. For each cluster $K \subset G^{n+1} \backslash G^{n+1,0}$, we define 
		\begin{equation}\label{defi-Omga-K-tau}
		\qquad \Omega_K = \bigcup_{Q \in K} Q,\quad \Omega_{K,\tau} = \bigcup_{Q \in K} Q_\tau,\quad Q_{\tau} = \{ x \in \R^d:  {\rm dist}(x,Q) <  \tau  |Q|^{1/d}  \}.
	\end{equation} 
\noindent
For each cluster $K$, we can construct an {\bf enlarged set} $\tilde \Omega_K $ of $\Omega_K$ such that $\Omega_{K,\frac{1}{4}} \subset \tilde \Omega_K \subset \Omega_{K,\frac{1}{2}}$, $\tilde \Omega_{K}$ is topologically connected,  and $\varepsilon^{-(n+1)}\tilde \Omega_K $ is $(\eta, L, M)$-uniformly of class  $C^{[s],1}$ as in Definition \ref{defi-class-C-m-1}. The detailed construction is given in Lemma \ref{lem-tilde-Omega-K} below. Note that $\tilde \Omega_{K_1} \cap \tilde \Omega_{K_2} = \emptyset$ for two different clusters $K_1,K_2$ (since $\Omega_{K_1,\frac 1 2} \cap \Omega_{K_2, \frac 1 2} = \emptyset$). Hence, we can define   $G^{n+1,1}$ as the collection of  all sub-cubes from all clusters $K \subset G^{n+1} \backslash G^{n+1,0}$ satisfying
	\begin{equation}\label{Condition-Q-in-G-n-1}
		\int_{\tilde \Omega_K}  \rho < 1+\delta.
	\end{equation}

	\item   $G^{n+1,2} $ is the collection of sub-cubes from clusters $K \subset G^{n+1} \backslash G^{n+1,0}$ satisfying
	\begin{equation}\label{Condition-Q-in-G-n-2}
		\int_{\tilde \Omega_K}  \rho \ge  1+\delta.
	\end{equation}	
\end{itemize}
\noindent
Since $0\le \rho \in L^1(\R^d)$, this procedure stops after finitely many steps and we obtain a collection of disjoint sub-cubes $\bigcup_{n\ge 1}  \{Q\}_{Q\in G^{n,0} \cup G^{n,1}}$ such that 
\begin{equation*}
	\text{supp}(\rho_\Psi) \subset \bigcup_{n \ge 1 } \left( \bigcup_{Q \in G^{n,0}\cup G^{n,1}} Q \right). 
\end{equation*}

 The above construction is similar to that in  \cite{KogNam-21}, except that here we have to construct the enlarged set $\tilde{\Omega}_K$ of $\Omega_K$ more carefully rather than using the simple choice $\Omega_{K,\frac 1 4}$ as in \cite{KogNam-21}. Heuristically, we want to choose $\tilde{\Omega}_K$ to be sufficiently smooth so that we have a better control of the localization error. To be precise, the construction of $\tilde{\Omega}_K$ is given by the following lemma. 
	
	\begin{lemma}[Construction of enlarged clusters]\label{lem-tilde-Omega-K}
		There exist $\eta, L, M>0$ depending only on $d,m\in \mathbb{N}$ such that  for any cluster $K$ in $G^{n+1} \backslash G^{n+1,0} $, there exists an enlarged set $\tilde \Omega_K $ of $\Omega_K$ such that $ \Omega_{K,\frac{1}{4}} \subset \tilde \Omega_K \subset \Omega_{K,\frac{1}{2}}$, $\tilde \Omega_{K}$ is topologically connected, and $\varepsilon^{-(n+1)}\tilde \Omega_K $ is  $(\eta, L, M)$-uniformly of class  $C^{m,1}$ as in Definition \ref{defi-class-C-m-1}. 
	\end{lemma}
	
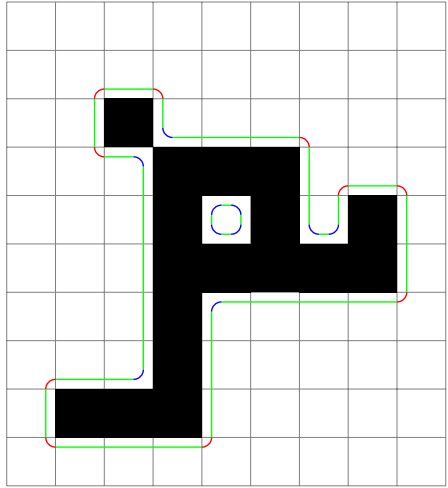
\begin{figure}[h]
	\centering
	\resizebox{0.4\textwidth}{!}{%
	\begin{tikzpicture}
	% Draw the grid
	\draw[step=1cm, gray, very thin] (0, 0) grid (9, 10);
	\fill[black] (1,1) rectangle (2,2);
	\fill[black] (2,2) rectangle (3,1);
	\fill[black] (3,1) rectangle (4,2);
	\fill[black] (3,2) rectangle (4,3);
	\fill[black] (3,3) rectangle (4,4);
	\fill[black] (3,4) rectangle (4,5);
	\fill[black] (3,5) rectangle (4,6);
	\fill[black] (4,4) rectangle (5,5);
	\fill[black] (5,4) rectangle (6,6);
	% Draw the black filled shape
	\fill[black] (3,6) rectangle (6,7);
	\fill[black] (6,4) rectangle (8,5);
	\fill[black] (7,5) rectangle (8,6);
	\fill[black] (3,7) rectangle (2,8);
	%red line
	\draw[ green, thick] (0.8,1) -- (0.8,2) ;
	\draw[ green, thick] (1,2.2) -- (2.6,2.2) ;
	\draw[ green, thick] (2.8, 2.4) -- (2.8, 6.6) ;
	\draw[ green, thick] (2.6, 6.8) -- (2, 6.8) ;
	\draw[ green, thick] (1.8, 7) -- (1.8, 8) ;
	\draw[ green, thick] (2, 8.2) -- (3, 8.2) ;
	\draw[ green, thick] (3.2, 8) -- (3.2, 7.4) ;
	\draw[ green, thick] (3.4, 7.2) -- (6, 7.2) ;
	\draw[ green, thick] (1, 0.8) -- (4, 0.8) ;
	\draw[ green, thick] (4.2, 1) -- (4.2, 3.6) ;
	\draw[ green, thick] (4.4, 3.8) -- (8, 3.8) ;
	\draw[ green, thick] (8.2, 4) -- (8.2, 6) ;
	
	\draw[ green, thick] (8,6.2) -- (7, 6.2) ;
	
	\draw[ green, thick] (6.8,6) -- (6.8, 5.4) ;
	
	\draw[ green, thick] (6.6,5.2) -- (6.4, 5.2) ;
	
	\draw[ green, thick] (6.2,5.4) -- (6.2, 7) ;

	\draw[ green, thick] (4.4,5.2) -- (4.6, 5.2) ;
	
	\draw[ green, thick] (4.4,5.8) -- (4.6, 5.8) ;
	
	\draw[green, thick] (4.2,5.4) -- (4.2, 5.6) ;
	
	\draw[green, thick] (4.8,5.4) -- (4.8, 5.6) ;

	\draw[blue, thick] (4.2,5.6) arc[start angle=180, end angle=90, radius=0.2cm];
	
	\draw[blue, thick] (4.2,5.4) arc[start angle=180, end angle=270, radius=0.2cm];
	
	\draw[blue, thick] (4.6,5.8) arc[start angle=90, end angle=0, radius=0.2cm];
	
	\draw[blue, thick] (4.8,5.4) arc[start angle=360, end angle=270, radius=0.2cm];
	%&\draw[blue, thick, domain=0.8:1, samples=20, smooth] 
	%\x)});
\draw[red, thick] (0.8,1) arc[start angle=180, end angle=270, radius=0.2cm];
\draw[red, thick] (1,2.2) arc[start angle=90, end angle=180, radius=0.2cm];
\draw[blue, thick] (2.6,2.2) arc[start angle=270, end angle=360, radius=0.2cm];

\draw[blue, thick] (2.6,6.8) arc[start angle=90, end angle=0, radius=0.2cm];

\draw[red, thick] (1.8,7) arc[start angle=180, end angle=270, radius=0.2cm];

\draw[red, thick] (1.8,8) arc[start angle=180, end angle=90, radius=0.2cm]; 

\draw[red, thick] (3,8.2) arc[start angle=90, end angle=0, radius=0.2cm];

\draw[blue, thick] (3.2,7.4) arc[start angle=180, end angle=270, radius=0.2cm];

\draw[red, thick] (6,7.2) arc[start angle=90, end angle=0, radius=0.2cm];

\draw[blue, thick] (6.2, 5.4) arc[start angle=180, end angle=270, radius=0.2cm];

\draw[blue, thick] (6.8, 5.4) arc[start angle=360, end angle=270, radius=0.2cm];

\draw[red, thick] (6.8, 6) arc[start angle=180, end angle=90, radius=0.2cm];

\draw[red, thick] (8, 6.2) arc[start angle=90, end angle=0, radius=0.2cm];

\draw[red, thick] (8.2, 4) arc[start angle=360, end angle=270, radius=0.2cm];

\draw[blue, thick] (4.4, 3.8) arc[start angle=90, end angle=180, radius=0.2cm];

\draw[red, thick] (4, 0.8) arc[start angle=270, end angle=360, radius=0.2cm];
\end{tikzpicture}	}
\caption{An enlarged set of a cluster $\Omega_K$ in 2D. The boundary of the enlarged set  is mostly flat, except at the corners where it must be smooth. In 2D, there are only two types of conners.}
	\label{fig:closure-2D}
\end{figure}

	\begin{proof} Since the case $d=1$ is trivial, let us focus on the case $d \ge 2$. Let $K \in G^{n+1}\backslash G^{n+1,0}$  be a cluster and let $\Omega_K$ be defined as in  \eqref{defi-Omga-K-tau}.  Then 
	\begin{align}\label{eq:K0-K}
	K_0=\epsilon^{-(n+1)} K=\{\epsilon^{-(n+1)} Q: Q \in K\}
	\end{align}
	is a cluster of unit cubes. We denote $ \Omega_{K_0} = \epsilon^{-(n+1)} \Omega_K$ and    
\begin{equation}\label{eq:def-K-tau}
	\Omega_{K_0,\frac{3}{16}} = \bigcup_{Q \in K_0} Q_{\frac{3}{16}} 
\end{equation}
where $Q_{\tau}$  was introduced  in \eqref{defi-Omga-K-tau}. We observe that  the boundary of  $\Omega_{K_0,\frac{3}{16}}$ is mostly flat, except a zero measure set  due to the combinations of connected unit cubes. Thus, the construction of the enlarged set $\tilde \Omega_{K_0}$ of $\Omega_{K_0}$ is essentially based on $\Omega_{K_0,\frac{3}{16}}$, plus smooth corrections  near the zero measure set that we will explain.

Let us consider $Q_{\frac{3}{16}}$ for a cube $Q \in K_0$. Then by the definition of $Q_{\tau}$, there exists $\vec J=(j_1,...,j_d) \in \Z^d$ such that 
$$ Q_{\frac{3}{16}} = \left\{ x \in \R^d: j_k -1 +\frac{3}{16} < x_k  < j_k + \frac{3}{16}, \forall k \in \{ 1,...,d\}  \right\}:= C_{\vec J}.$$ 
Therefore,  
\begin{align*}
	\partial Q_{\frac{3}{16}} &= \partial C_{\vec J} = \bar C_{\vec J} \backslash C_{\vec J} = \bigcup_{k=1}^d   S^+_{\vec J,k} \cup S^-_{\vec J,k},
\end{align*}
where  \begin{align*}
	S_{\vec J,k}^{\pm} &= \left\{ x=(x_1,...,x_d) \in  \partial C_{\vec J}: x_k = j_k \pm \frac{3}{16}         \right\}
	\end{align*}
which are interpreted as a modification of the ``hyper-surfaces'' of the cube.  Now, let us construct open covers for $S_{\vec J,k}^+ $. For $\beta \in (0,1)$ small enough, we introduce the covering 
\begin{align*}
	S_{\vec J,k}^{\pm} \subset  B_{\vec J, k, 0}^{\pm} \cup \left(\bigcup_{\substack{m,n \ge 0\\m+n=d-1}} B^{\pm}_{\vec J,k;k_1,...,k_m;\ell_1,...,\ell_n } \right)  \end{align*}
where
\begin{align*}
	B_{\vec J, k, 0}^+ = \left\{ x \in \R^d:  \left|x_k - (j_k + \frac{3}{16})\right| < \beta, j_i - \frac{13}{16} +\frac{\beta}{4} <  x_i < j_i + \frac{3}{16} -\frac{\beta}{4}, \forall i \ne k   \right\},\\
	B_{\vec J, k, 0}^- = \left\{ x \in \R^d:  \left|x_k - (j_k - \frac{13}{16})\right| < \beta, j_i - \frac{13}{16} +\frac{\beta}{4} <  x_i < j_i + \frac{3}{16} -\frac{\beta}{4}, \forall i \ne k   \right\},
\end{align*}	
and for any partition of $\{1,...,d\} = \{k,k_1,...,k_m,\ell_1,...,\ell_n\}$, 
\begin{align*}
	&B^+_{\vec J, k; k_1,...,k_m;\ell_1,...,\ell_n}\\
	&=\left\{ x \in \R^d:  \left|x_k - (j_k + \frac{3}{16})\right| < \beta,    \left|x_{\ell_i} - (j_{\ell_i} + \frac{3}{16})\right| < \beta, \left|x_{k_q} - (j_{k_q} - \frac{13}{16})\right| < \beta    \quad  \forall i, q    \right\}\\
	& B^-_{\vec J, k; k_1,...,k_m;\ell_1,...,\ell_n}\\
	&=\left\{ x \in \R^d:  \left|x_k - (j_k - \frac{13}{16})\right| < \beta,    \left|x_{\ell_i} - (j_{\ell_i} + \frac{3}{16})\right| < \beta, \left|x_{k_q} - (j_{k_q} - \frac{13}{16})\right| < \beta    \quad  \forall i, q    \right\}.
\end{align*}	
Thus, the collection $\{B_{\vec J, k, 0}^{\pm},B^{\pm}_{\vec J, k; k_1,...,k_m;\ell_1,...,\ell_n}\}$ are an open covering for the boundary  $\partial Q$.

In particular, note that $\partial Q \cap B_{\vec J, k, 0}^{\pm} $ is flat and   there exist  continuous functions $f^{\pm}_{\vec J, k; k_1,...,k_m;\ell_1,...,\ell_n}$  such that 
$$ Q \cap B^{\pm}_{\vec J, k; k_1,...,k_m;\ell_1,...,\ell_n} = B^{\pm}_{\vec J, k; k_1,...,k_m;\ell_1,...,\ell_n} \cap \{ x_N > f^{\pm}(x_1,...,x_N)  \}. $$      
In particular, the  number of the family $ \{f^{\pm}\}_{ k; k_1,...,k_m;\ell_1,...,\ell_n}$ is finite, depending only on the dimension $d$. We can further approximate $ \{f^{\pm}\}$ by smooth functions $\{ \tilde f^{\pm} \}$ such that 
$$  \|\tilde f^{\pm } - f^{\pm}\|_{L^\infty} \le \eps_d \beta    $$
where the small constant $\eps_d>0$ can be chosen depending only on $d$ such that the set 
$$  \tilde Q = \bigcup B_{\vec J, k, 0}^{\pm} \bigcup B^{\pm}_{\vec J, k; k_1,...,k_m;\ell_1,...,\ell_n} \cap \{ x_N > \tilde f^{\pm}(x_1,...,x_N)  \}$$
satisfies
$$  Q_{\frac{1}{8}}  \subset \tilde Q \subset Q_{\frac{1}{4}}.$$

In this way, we have obtained $\tilde Q$ as an enlarged set of each cube $Q\in K_0$ such that the boundary $\partial \tilde Q$ is $(\eta, L, M)$-uniformly of class $C^{m,1}$  as in Definition \ref{defi-class-C-m-1}. Since the cluster $K_0$ is constructed from unit cubes, and all unit cubes are identical up to translations, we can apply the same enlarging procedure as above. Note that each cube in $K_0$ can be only connected to at most $\tilde C(d)$  cubes, where $\tilde C(d)$ depends only on $d$, by repeating the previous construction we will only need to use a finitely many types  of $\tilde f^{\pm}$, where the number depends only on $d$. 

By scaling, the construction for $K_0$ leads to the desired construction for $K$. The proof of Lemma \ref{lem-tilde-Omega-K} is complete. For an illustration of the construction in two dimensions, the reader may see Figure \ref{fig:closure-2D}
\end{proof}

\iffalse 
\begin{figure}[h]
			\centering
			\includegraphics[width=0.5\textwidth]{square}
			\caption{An enlarged set of a cluster $\Omega_K$ in 2D. The boundary of the enlarged set  is mostly flat, except at the corners where it must be smooth. In 2D, there are only two types of conners.}
		%	\label{fig:closure-2D}
		\end{figure}%picture
\fi

\section{Local uncertainty estimates}\label{section-local-uncertainty-principles} 

%We will prove the main results in Theorem \ref{thm:1-fractional} and Theorem \ref{thm:2-fractional} using microlocal techniques. The proofs require suitable local versions of the uncertainty and exclusion principles, which will be derived in this section. 
We assume that $\Psi$ is a normalized wave function in $L^2(\R^{dN})$ such that $\Psi\in H^s(\R^{dN})$ and its one-body density $\rho_\Psi$ is supported in $[-1/2,1/2]^{d}$. As usual, we write $s = m +\sigma>0$ with  $m = [s]$ and $\sigma \in [0,1)$.  We construct the covering sub-cubes of $[-1/2,1/2]^{d}$ as in Section \ref{section-subcovering}, using $\rho=\rho_\Psi$ and two parameters $\delta \in (0,1)$ and    
$\varepsilon \in  \{\frac{1}{2}, \frac{1}{3}\}$. 

 The following result will be needed for the proof of Theorem \ref{thm:1-fractional}.

\begin{lemma}[Local uncertainty principles]\label{lemma-local-uncertainty-II} Let $K \in G^{n}\backslash G^{n,0}$  be a cluster which contains $|\#K|$ sub-cubes of volumes $\eps^{n}$. Let  $\Omega_{K}$  be as in \eqref{defi-Omga-K-tau} and let $\tilde \Omega_{K}$ be the enlarged set of $\Omega_{K}$ constructed in Lemma \ref{lem-tilde-Omega-K}. Then for all $\xi\in (0,1)$ we have
	\begin{align}
		\left\langle \Psi, \left( \sum_{i=1}^N (-\Delta_{x_i})^s_{| \tilde \Omega_{K}} \right) \Psi\right\rangle &\ge  C_{ \rm GN} (1 - \xi)\frac{\int_{\Omega_{K}}  \rho_{\Psi}^{1+\frac{2s}{d}}}{\left(\int_{ \tilde \Omega_{K}}  \rho_{\Psi} \right)^\frac{2s}{d}} \nonumber\\
		&\qquad - \frac{C(d,s)(1+ |\# K|^{\frac{2\sigma}{d}})^{\frac{s}{s-t_0}}}{\xi^{\frac{t_0}{s-t_0}} \eps^{2sn}}\int_{ \tilde \Omega_{K}} \rho_{\Psi} \label{esti-local-uncertainty-II}
	\end{align} 
	where  $C_{\rm GN}$ is the Gagliardo--Nirenberg constant defined in \eqref{eq:GN-constant-s} and 
	\begin{equation}\label{defi-t-0}
		t_0 = \begin{cases}
			 s  -  1   \quad &\text{ if } \sigma(s) = 0, \\
			  s - \frac{\min(\sigma(s), 1 -\sigma(s))}{4} \quad  &\text{ if } \sigma(s) \in (0,1).
		 \end{cases}
	\end{equation}
	In particular, for each cube $Q \in K$, one has
	\begin{equation}\label{uncertainty-type-I}
		\left\langle \Psi, \left( \sum_{i=1}^N (-\Delta_{x_i})^s_{| Q} \right) \Psi\right\rangle   \ge \frac{1}{C(d,s)} \dfrac{\int_Q \rho_\Psi^{1+ \frac{2s}{d}}}{ (\int_Q \rho_\Psi)^{\frac{2s}{d}}} - \frac{C(d,s)}{|Q|^{\frac {2s} d}} \int_Q \rho_\Psi. 
	\end{equation}
\end{lemma}
\begin{proof} 
The proof  of  \eqref{uncertainty-type-I} is the same 
	\cite[Lemma 3]{KogNam-21} (see also \cite[Lemma 8]{LuNaPo-16} and \cite[Lemma 3]{Nam-22}). Hence, we only  prove  \eqref{esti-local-uncertainty-II} here. By the argument in  \cite[p. 1180]{KogNam-21},  the inequality  \eqref{esti-local-uncertainty-II}   follows from the one-body estimate 
	\begin{equation}\label{esti-nbabla-u-ge-wit-u-in-H-1}
		\|  u\|^2_{\dot H^{s}(\tilde \Omega_{K})} \ge C_{\rm GN} (1 - \xi)\frac{\int_{\Omega_{K}}   |u|^{2\left(1+\frac{2s}{d} \right)}}{\left(\int_{\tilde \Omega_{K}}  |u|^2 \right)^\frac{2s}{d}} - \frac{C(d,s)(1+|\#K|^\frac{2\sigma}{d})^{\frac{s}{s-t_0}}}{ \xi^{\frac{t_0}{s-t_0}} \eps^{2sn}} \|u\|^2_{L^2(\tilde \Omega_{K})},
	\end{equation} 
	for all $u \in H^s(\R^d)$. We will prove  \eqref{esti-nbabla-u-ge-wit-u-in-H-1} using the fact that $\varepsilon^{-n}\tilde \Omega_K $ is  $(\eta, L, M)$  uniformly of class  $C^{m,1}$, which is due to Lemma \ref{lem-tilde-Omega-K}. By a simple scaling argument, it suffices to prove \eqref{esti-nbabla-u-ge-wit-u-in-H-1} where $K$ is a collection of unit cubes and $\eps^{2ns}$ is replaced by $1$. 
%	, \xi \in (0,1)$ and  $t_0$ is defined by
%	\begin{equation}\label{defi-t-0}
%		t_0 = \left\{ 	\begin{array}{rcl}
%			&&	s  -  1   \text{ if } \sigma(s) =0,\\[0.3cm]
%			& &  s - \frac{\min(\sigma(s), 1 -\sigma(s))}{4},  \text{ if } \sigma(s) \in (0,1).
%		\end{array}\right.
%	\end{equation}   
	
	\smallskip

	Let $ \chi,\eta: \R^d \to [0,1] $ be smooth functions satisfying 
	\begin{equation}\label{defi-chi-eta}
		\chi^2 +   \eta^2 =1, \quad  \chi (x) =1 \text{ for all } x \in \Omega_{K},\quad   \text{supp}(\chi) \subset  \Omega_{K, \frac{1}{16}} \subset \tilde{\Omega}_K,
	\end{equation}
	and   
	\begin{equation}\label{dist-supp-chi-tilde-Omega-c}
		\zeta = \text{ \rm dist}({\rm supp}(\chi), \tilde \Omega^c_{K}) \ge \frac{1}{16}, \quad  \|\eta\|_{W^{m+1,\infty}(\R^d)} + \|\chi\|_{W^{m+1,\infty}(\R^d)} \lesssim_{d,m} 1.
	\end{equation}
	Here $\Omega_{K, \frac{1}{16}}$ is defined similarly to \eqref{eq:def-K-tau}. 	By the Gagliardo--Nirenberg inequality,  
	\begin{equation}\label{GNinequa-for-chi-u}
		\|\chi u\|_{\dot{H}^s(\R^d)}^2 \ge C_{\rm GN} \frac{\int_{\R^d} |\chi u|^{2(1+ \frac{2s}{d})} }{\left( \int_{\R^d} |\chi u|^2 \right)^\frac{2s}{d} } \ge C_{\rm GN}  \frac{\int_{\Omega_{K}} |u|^{2(1+ \frac{2s}{d})} }{\left( \int_{ \tilde \Omega_{K}} | u|^2 \right)^\frac{2s}{d} }.
	\end{equation} 
In the following, we will prove that
	\begin{equation}\label{es-chi-u-control-chi u tilde Omega}
		\|\chi u\|^2_{\dot{H}^s(\R^d)}	 \le  (1 + \xi) \|\chi u\|^2_{\dot{H}^s(\tilde \Omega_{K})} +  \frac{C(d,s)}{ \xi^{\frac{t_0}{s -t_0}}} \|u\|^2_{L^2(\tilde \Omega_{K})},
	\end{equation}
	and 
	\begin{equation}\label{esti-chi-u-control-by-u}
		\|\chi u\|_{\dot{H}^s(\widetilde{\Omega}_{K})}^2 \leq(1+\xi)\|u\|_{\dot{H}^s({\tilde \Omega}_{K})}^2 + \frac{C(d,s)(1+|\# K|^\frac{2\sigma}{d})^{\frac{s}{s-t_0}}}{  \xi^{\frac{t_0}{s-t_0}}  } \|u\|_{L^2(\widetilde{\Omega}_{K})}^2.
	\end{equation}
	\noindent 
	Then  the desired estimate  \eqref{esti-nbabla-u-ge-wit-u-in-H-1} follows immediately from \eqref{GNinequa-for-chi-u}, \eqref{es-chi-u-control-chi u tilde Omega} and \eqref{esti-chi-u-control-by-u}.

	\smallskip
	\noindent 
	\textbf{Proof of \eqref{es-chi-u-control-chi u tilde Omega}:}   We denote $\tilde \Omega = \tilde \Omega_{K}$ for simplicity. We observe that  if $s =m \in \N$, then $\|\chi u\|^2_{\dot H^m(\R^d)}  = \|\chi u\|^2_{\dot H^m(\tilde \Omega)}$
	which yields \eqref{es-chi-u-control-chi u tilde Omega} without  the term involving $\|u\|^2_{L^2(\tilde \Omega)}$. For  $s \notin \N$, from \eqref{defi-seminorm-dot-H-s} it follows that 
	$$\begin{aligned} \|\chi u\|_{\dot{H}^s\left(\mathbb{R}^d\right)}^2 &=\|\chi u\|_{\dot{H}^s(\widetilde{\Omega})}^2+\sum_{|\alpha|=m} 2 c_{d, \sigma} \frac{m !}{\alpha !} \int_{\widetilde{\Omega}} \int_{\mathbb{R}^d \backslash \widetilde{\Omega}} \frac{\left|\partial^\alpha(\chi u)(x)-\partial^\alpha(\chi u)(y)\right|^2}{|x-y|^{d+2 \sigma}} d x d y \\ &=\|\chi u\|_{\dot{H}^s(\widetilde{\Omega})}^2+\sum_{|\alpha|=m} 2 c_{d, \sigma} \frac{m !}{\alpha !} \int_{\widetilde{\Omega}} \int_{\mathbb{R}^d \backslash \widetilde{\Omega}} \frac{\left|\partial^\alpha(\chi u)(y)\right|^2}{|x-y|^{d+2 \sigma}} d x d y .
	\end{aligned}$$
	Using    $ \zeta =   {\rm dist}({\rm supp}(\chi), \tilde \Omega^c) \ge \frac{1}{16}$,   we derive  from  \eqref{dist-supp-chi-tilde-Omega-c} that  
	\begin{align}\label{compa-chi-u-R-d-chi-u-tilde-Omega-u-H-m}
			\|\chi u\|_{\dot{H}^s\left(\mathbb{R}^d\right)}^2& \leq\|\chi u\|_{\dot{H}^s(\widetilde{\Omega})}^2+\sum_{|\alpha|=m} 2 c_{d, \sigma} \frac{m !}{\alpha !} \int_{\widetilde{\Omega}}\left|\partial^\alpha(\chi u)(y)\right|^2 d y \int_{|x| \ge \frac{\zeta}{2}} \frac{1}{|x|^{d+2 \sigma}} d x \nonumber\\ & \le \|\chi u\|_{\dot{H}^s(\widetilde{\Omega})}^2 + C(d,s)\|\chi u\|_{\dot{H}^m(\widetilde{\Omega})}^2.  
	\end{align}
	Besides that, we can use \eqref{dot-H-t-dot-H-s} and Young's inequality which give 
	\begin{equation*}
		\|\chi u\|_{\dot{H}^m(\widetilde{\Omega})}^2 = \|\chi u\|_{\dot{H}^m(\R^d)}^2  \le  a \| \chi u\|^2_{\dot H^s(\R^d)} + \frac{C(d, s)}{a^\frac{m}{s-m}} \| \chi u\|^2_{L^2(\R^d)}
	\end{equation*}
	for all $a>0$. 	Hence, we derive from \eqref{compa-chi-u-R-d-chi-u-tilde-Omega-u-H-m} that
	\begin{equation*}
		\begin{aligned} 
			\|\chi u\|_{\dot{H}^s\left(\mathbb{R}^d\right)}^2 & \le \frac{1}{1-C(d,s)a} \|\chi u\|_{\dot{H}^s(\widetilde{\Omega})}^2 + \frac{C(d, s)}{a^\frac{m}{s-m}} \|\chi u\|_{\dot{H}^m(\widetilde{\Omega})}^2.  \end{aligned}
	\end{equation*}	
	Finally, by taking $\xi = \frac{C a}{1 -Ca}$ with $a>0 $ small enough and using 
	$$\frac{m}{s-m} \le \frac{t_0}{s-t_0}$$
	which comes from the definition of $t_0$ in \eqref{defi-t-0}, we conclude \eqref{es-chi-u-control-chi u tilde Omega}.
	
	\smallskip 
	
	\noindent
	\textbf{Proof of  \eqref{esti-chi-u-control-by-u}:} Let us start by proving that the following result from \cite[Lemma 14]{LuNaPo-16}, 
	\begin{align}
		&\left|\|u\|_{\dot{H}^s(\tilde \Omega)}^2-\|\chi u\|_{\dot{H}^s(\tilde \Omega)}^2-\|\eta u\|_{\dot{H}^s(\tilde \Omega)}^2\right| 
		\le C(d,s)(1 +|\tilde \Omega|^{\frac{2\sigma}{d}})\left(\|\chi u\|_{H^{t_0}(\tilde \Omega)}^2+\|\eta u\|_{H^{t_0}(\tilde \Omega)}^2\right),\label{norm-u-chi-eta-u-le-t-0-notinteger}
	\end{align} 	
	holds with the explicit choice of $t_0$ in \eqref{defi-t-0}. 
	
	We will distinguish again the case $s\in \mathbb{N}$ and the case  $s\notin \mathbb{N}$. For  $s = m \in \N$,   we have
	\begin{align*}
		& \left|D^{\alpha}(\chi u)\right|^{2}+\left|D^{\alpha}(\eta u)\right|^{2} - \left(\chi^{2}+\eta^{2}\right)\left|D^{\alpha} u\right|^{2} \\
		&= 2 \Re \sum_{\beta<\alpha} \frac{\alpha!}{\beta!(\alpha-\beta)!}\left(\chi D^{\alpha-\beta} \chi+\eta D^{\alpha-\beta} \eta\right) D^{\alpha} \bar{u} D^{\beta} u \\ 
		& \quad +\left|\sum_{\beta<\alpha} \frac{\alpha!}{\beta!(\alpha-\beta)!} D^{\alpha-\beta} \chi D^{\beta} u\right|^{2}+\left|\sum_{\beta<\alpha} \frac{\alpha!}{\beta!(\alpha-\beta)!} D^{\alpha-\beta} \eta D^{\beta} u\right|^{2}
	\end{align*}
	with any multi-index $\alpha$  satisfying $|\alpha| =m$. 	By the Cauchy-Schwarz inequality,  
	$$
	J_1:=\left|\sum_{\beta<\alpha} \frac{\alpha!}{\beta!(\alpha-\beta)!} D^{\alpha-\beta} \chi D^{\beta} u\right|^{2}+\left|\sum_{\beta<\alpha} \frac{\alpha!}{\beta!(\alpha-\beta)!} D^{\alpha-\beta} \eta D^{\beta} u\right|^{2} \le C(d,s) \sum_{\beta<\alpha}\left|D^{\beta} u\right|^{2},
	$$
	which yields
	$$ \int_{\tilde \Omega}  J_1 \le C(d,s) \|u\|^2_{H^{m-1}(\tilde \Omega)}.$$
	Moreover, if we define 
	$$J_2 = 2 \Re \sum_{\beta<\alpha} \frac{\alpha!}{\beta!(\alpha-\beta)!}\left(\chi D^{\alpha-\beta} \chi+\eta D^{\alpha-\beta} \eta\right) D^{\alpha} \bar{u} D^{\beta} u, $$
	then by using   integration by parts in combining with \eqref{dist-supp-chi-tilde-Omega-c}, we deduce that 
	$$ \int_{\tilde \Omega}  |J_2|  \le C(d,s) \|u\|^2_{H^{m-1}(\tilde \Omega)}.   $$
	By adding the above estimates for $J_1$ and $J_2$,  we obtain the desired bound \eqref{norm-u-chi-eta-u-le-t-0-notinteger}.

	\smallskip 
	Next, we consider the case  $s \notin  \N$, namely $\sigma =s-m \in (0,1)$. Define   $\omega = \frac{1}{2} \min(\sigma, 1-\sigma) $. Using \eqref{dist-supp-chi-tilde-Omega-c}, we have
		\begin{align*}
		& \left|\left|D^{\alpha}(\chi u)(x)-D^{\alpha}(\chi u)(y)\right|^{2}-\left|\chi(x) D^{\alpha} u(x)-\chi(y) D^{\alpha} u(y)\right|^{2}\right| \\
		& \quad \le C|x-y|^{2 \omega}\left|D^{\alpha}(\chi u)(x)-D^{\alpha}(\chi u)(y)\right|^{2} \\
		& \quad\quad +C \sum_{\beta<\alpha}\left(1+|x-y|^{-2 \omega}\right)\left(\left|D^{\beta} u(x)-D^{\beta} u(y)\right|^{2}+|x-y|^{2}\left|D^{\beta} u(y)\right|^{2}\right) 
		\end{align*}
		and 
		\begin{align*}
		& \left|\left|D^{\alpha}(\eta u)(x)-D^{\alpha}(\eta u)(y)\right|^{2}-\left|\eta(x) D^{\alpha} u(x)-\eta(y) D^{\alpha} u(y)\right|^{2}\right| \\
		& \quad \le C|x-y|^{2 \omega}\left|D^{\alpha}(\eta u)(x)-D^{\alpha}(\eta u)(y)\right|^{2} \\
		& \quad\quad +C \sum_{\beta<\alpha}\left(1+|x-y|^{-2 \omega}\right)\left(\left|D^{\beta} u(x)-D^{\beta} u(y)\right|^{2}+|x-y|^{2}\left|D^{\beta} u(y)\right|^{2}\right)  
	\end{align*}
	and
	\begin{align*}
		& \left|\left|\chi(x) D^{\alpha} u(x)-\chi(y) D^{\alpha} u(y)\right|^{2}+\left|\eta(x) D^{\alpha} u(x)-\eta(y) D^{\alpha} u(y)\right|^{2} - \left|D^{\alpha} u(x)-D^{\alpha} u(y)\right|^{2} \right| \\
		& \quad=\left|\left((\chi(x)-\chi(y))^{2}+(\eta(x)-\eta(y))^{2}\right) \Re D^{\alpha} \bar{u}(x) D^{\alpha} u(y)\right| \\
		& \quad \leq C|x-y|^{2}\left(\left|D^{\alpha} u(x)\right|^{2}+\left|D^{\alpha} u(y)\right|^{2}\right) ,
	\end{align*}
with a universal constant $C =C(d,s)>0$. Then,  by the triangle inequality,   
	$$
	\begin{aligned}
		& \left\lvert\, \iint_{\tilde \Omega \times \tilde \Omega} \frac{\left|D^{\alpha}(\chi u)(x)-D^{\alpha}(\chi u)(y)\right|^{2}+\left|D^{\alpha}(\eta u)(x)-D^{\alpha}(\eta u)(y)\right|^{2}}{|x-y|^{d+2 \sigma}} \mathrm{d}x \mathrm{d}y\right. \\
		& \left.\quad-\iint_{\tilde \Omega \times \tilde \Omega} \frac{\left|D^{\alpha} u(x)-D^{\alpha} u(y)\right|^{2}}{|x-y|^{d+2 \sigma}} \mathrm{d} x \mathrm{d} y \right\rvert\,  \quad \leq C(d,s) ( I_1 + I_2 +I_3),
	\end{aligned}
	$$  
	where
	\begin{align*}
		I_1 &= \iint_{\tilde \Omega \times \tilde \Omega} \frac{\left|D^{\alpha}(\chi u)(x)-D^{\alpha}(\chi u)(y)\right|^{2}}{|x-y|^{d+2(\sigma- \omega)}} dxdy + \iint_{\tilde \Omega \times \tilde \Omega} \frac{\left|D^{\alpha}(\eta u)(x)-D^{\alpha}(\eta u)(y)\right|^{2}}{|x-y|^{d+2(\sigma- \omega)}} dxdy ,  \\
		I_2 & =    \sum_{\beta<\alpha} \iint_{\tilde \Omega \times \tilde \Omega} \frac{\left(1+|x-y|^{-2 \omega}\right)}{|x-y|^{d+2\sigma}}\left(\left|D^{\beta} u(x)-D^{\beta} u(y)\right|^{2}+|x-y|^{2}\left|D^{\beta} u(y)\right|^{2}\right)dx dy , \\
		I_3  &=  \iint_{\tilde \Omega \times \tilde \Omega} \frac{\left|D^{\alpha} u(x)\right|^{2} +\left|D^{\alpha} u(y)\right|^{2}}{|x-y|^{d-2+2\sigma}}  dxdy .
	\end{align*}
	According to the definition in \eqref{defi-seminorm-dot-H-s}, we can estimate  
	$$ I_1 \le C(d,s)\left( \left\|D^{\alpha}(\chi u)\right\|_{\dot{H}^{\sigma-\omega}(\tilde \Omega)}^{2} + \left\|D^{\alpha}(\eta u)\right\|_{\dot{H}^{\sigma-\omega}(\tilde \Omega)}^{2}\right).$$
	Moreover, by  Fubini's theorem and the rearrangement inequality  \cite[Theorem 3.4]{LiLo-01}
	$$ \int_{\tilde \Omega}\frac{1}{|x-y|^{d-2+2\sigma}} dy \le C|\tilde \Omega|^{\frac{2\sigma}{d}}, $$
 we obtain
	\begin{align*}
		I_3 \le 2 \int_{\tilde \Omega} |D^\alpha u(x)|^2 \left(\int_{\tilde \Omega}\frac{1}{|x-y|^{d-2+2\sigma}} dy \right) dx \le C(d,s)  |\tilde \Omega|^{\frac{2\sigma}{d}} \|u\|^2_{\dot H^m(\tilde \Omega)}.
	\end{align*}
		Similarly, 
	\begin{align*}
		&\sum_{\beta<\alpha} \iint_{\tilde \Omega \times \tilde \Omega} \frac{\left(1+|x-y|^{-2 \omega}\right)}{|x-y|^{d+2\sigma-2}}\left|D^{\beta} u(y)\right|^{2}dx dy \\
		&\le C(d,s)  |\tilde \Omega|^{\frac{2\sigma}{d}} \left( \|u\|^2_{ H^{m-1}(\tilde \Omega)}\1_{s>1} + \|u\|^2_{H^{\sigma +\omega}(\tilde \Omega)}  \right).
	\end{align*}
	Moreover, by the definition in \eqref{defi-seminorm-dot-H-s}, for $|\beta|  < |\alpha| =m$, we have
	\begin{align*}
		&\sum_{\beta<\alpha} \iint_{\tilde \Omega \times \tilde \Omega} \frac{\left(1+|x-y|^{-2 \omega}\right)}{|x-y|^{d+2\sigma}}\left(\left|D^{\beta} u(x)-D^{\beta} u(y)\right|^{2}\right)dx dy\\
		& \le C(d,s)  \left( \|u\|^2_{ H^{m-1}(\tilde \Omega)}\1_{s>1} + \|u\|^2_{H^{\sigma +\omega}(\tilde \Omega)}  \right).
	\end{align*}
	By taking the sum over multi-indexes $\alpha$ satisfying $|\alpha| =m$, we arrive at 
	\begin{align*}
		& \left|\|\chi u\|_{\dot{H}^{s}(\tilde \Omega)}^{2}+\|\eta u\|_{\dot{H}^{s}(\tilde \Omega)}^{2}-\|u\|_{\dot{H}^{s}(\tilde \Omega)}^{2}\right| 
		\le C(d,s)(1 + |\tilde \Omega|^\frac{2\sigma}{d})\left( \|\chi u\|_{{H}^{s-\omega}(\tilde \Omega)}^{2}+\|\eta u\|_{{H}^{s-\omega}(\tilde \Omega)}^{2}  \right) .
	\end{align*}
	Thus  \eqref{norm-u-chi-eta-u-le-t-0-notinteger} holds for all $s>0$. The right-hand side of	 \eqref{norm-u-chi-eta-u-le-t-0-notinteger} can be estimated further using \eqref{compa-H-t-dot-H-s-impro} for $t =t_0$, we get 
	\begin{align*}
		&\left|\|u\|_{\dot{H}^s(\tilde \Omega)}^2-\|\chi u\|_{\dot{H}^s(\tilde \Omega)}^2-\|\eta u\|_{\dot{H}^s(\tilde \Omega)}^2\right| \\
		& \le  C  (1 + |\tilde \Omega|^\frac{2\sigma}{d})\tilde \xi \left(\|\chi u\|_{\dot H^{s}(\tilde \Omega)}^2+\|\eta u\|_{\dot H^{s}(\tilde \Omega)}^2\right)   +   \frac{C(1 + |\tilde \Omega|^\frac{2\sigma}{d})}{\tilde \xi^{\frac{t_0}{s-t_0}}} \|u\|^2_{L^2(\tilde \Omega)}
	\end{align*}
	for all $\tilde \xi>0$. 	Taking $\xi = C(d,s)(1 + |\tilde \Omega|^\frac{2\sigma}{d})\tilde \xi$, we obtain   
	\eqref{esti-chi-u-control-by-u}. The proof of Lemma \ref{lemma-local-uncertainty-II} is complete. 
\end{proof}

Next, we extend the above local uncertainty principle for  the  Hardy--Schr\"odinger operator. The following result will be helpful for the proof of Theorem \ref{thm:2-fractional}. 

\begin{lemma}\label{lemma-for-Uncertainty} Let $0<s<d/2$. Under the same assumptions in Lemma \ref{lemma-local-uncertainty-II}, we have for every $\xi \in (0,1)$,   
	\begin{equation}\label{uncertainty-Type-II-Hardy-operator}
		\left\langle \Psi,   \sum_{i=1}^N \left( (-\Delta)^s_{|\tilde \Omega_{K}} - \frac{\mathcal{C}_{d,s}}{|x_i|^{2s}} \1_{\Omega_{K}} \right) \Psi \right\rangle
		\ge  C_{\rm HGN} (1 - \xi)  \frac{\int_{\Omega_{K}}  \rho_{\Psi}^{(1+\frac{2s}{d})}}{\left(\int_{\tilde \Omega_{K}}  \rho_\Psi \right)^\frac{2s}{d}} 
		- \frac{\mathcal{H}(d,s,\xi,K)}{\eps^{2ns}}\int_{\tilde \Omega_{K}} \rho_{\Psi},
	\end{equation}
	where $C_{\rm HGN}$ is the with the Hardy--Gagliardo--Nirenberg constant in \eqref{constant-HGN} and 
	$$  \left|\mathcal{H}(d,s,\xi,K) \right| \le C(d,s) \left\{  \frac{1}{ \xi^{\frac{s+t_1}{s-t_1}}}  + (1 +|\# K|^\frac{2\sigma}{d})^\frac{t_1}{t_1-t_0} \right\},$$ 
	with $t_0$ given in \eqref{defi-t-0} and 
	\begin{align}\label{defi-t-1}
		t_1  =  
		\begin{cases} 
		 s-1    \quad & \text{ if } \sigma(s) = 0,  \\
			 s - \frac{\min(\sigma(s),1 - \sigma(s))}{8} \quad &\text{ if } \sigma(s) > 0. 
		\end{cases}
	\end{align}
	In particular, if $Q$ is a cube in $\R^d$ centered at $0$, then we  have 
	\begin{equation}\label{uncertainty-I-hardy-type-ope}
		\left\langle \Psi,   \sum_{i=1}^N \left( (-\Delta)^s_{|Q} - \frac{\mathcal{C}_{d,s}}{|x_i|^{2s}} \1_{Q}(x_i) \right) \Psi \right\rangle \ge  \frac{1}{C(d,s)} \dfrac{\int_Q \rho_\Psi^{1+ \frac{2s}{d}}}{ (\int_Q \rho_\Psi )^{\frac{2s}{d}}} - \frac{C(d,s)}{|Q|^\frac{2s}{d}} \int_Q \rho_\Psi.  
	\end{equation}
\end{lemma}
\begin{proof} We refer to  \cite[Eq. 33]{KogNam-21}  for the proof of \eqref{uncertainty-I-hardy-type-ope}.  Now, we prove \eqref{uncertainty-Type-II-Hardy-operator}. By the same explanation in the proof of Lemma  \ref{lemma-local-uncertainty-II}, it is sufficient to prove

	\begin{align}
			\|u\|^2_{\dot H^s(\tilde \Omega)}  - \mathcal{C}_{d,s}\int_\Omega \frac{|u(x)|^2}{|x|^{2s}} dx &\ge C_{\rm HGN}  (1 - \xi)\frac{\int_{\Omega}  \rho_{\Psi}^{1+\frac{2s}{d}}}{\left(\int_{\tilde \Omega}  \rho_{\Psi} \right)^\frac{2s}{d}} \noindent \\
			 &- \frac{C(d,s)}{\varepsilon^{2sn}} \left\{  \frac{1}{ \xi^{\frac{s+t_1}{s-t_1}}}  + (1 +|\# K|^\frac{2\sigma}{d})^\frac{t_1}{t_1-t_0} \right\}\int_{\tilde \Omega} \rho_{\Psi},\label{one-body-hardy-unvertainty}
	\end{align}
	where $t_0$ and $t_1$ defined by  \eqref{defi-t-0} and \eqref{defi-t-1}, respectively.  In addition, by a simple scaling argument, it suffices to prove \eqref{one-body-hardy-unvertainty} where $K$ is a collection of unit cubes and $\eps^{2ns}$ is replaced by $1$.

	\smallskip 

	Let $\chi $ and $\eta$   defined as in \eqref{defi-chi-eta}. By Hardy Gagliardo Nirenberg inequality, we have 
	\begin{equation}\label{HGN-inequality}
		\| \chi u\|^2_{\dot H^s(\R^d)} - \mathcal{C}_{d,s}\int_\Omega \frac{|\chi u(x)|^2}{|x|^{2s}} dx  \ge   C_{\rm HGN}  \frac{\int_{\R^d} |\chi u|^{2(1+ \frac{2s}{d})} }{\left( \int_{\R^d} |\chi u|^2 \right)^\frac{2s}{d} } \ge C_{\rm HGN}  \frac{\int_{\Omega} |u|^{2(1+ \frac{2s}{d})} }{\left( \int_{\tilde \Omega} | u|^2 \right)^\frac{2s}{d} }. 
	\end{equation}
	According to  \cite[Eq. (1.8)]{Frank-09}, 	one has 
	\begin{equation*}
		(-\Delta)^s-\mathcal{C}_{s, d}|x|^{-2s} \geq \ell^{(s-t)}(-\Delta)^t -  C(d,s,t)  \ell^{s},  \text{ on } L^2(\R^d),
	\end{equation*}
	for each $ 0 < s < \frac{d}{2}, 0 < t < s,$ and $ \ell >0 $. By taking $ \ell = \xi^{-\frac{2}{s-t}},$ we obtain
	\begin{equation*}
		\xi\left(\|\chi u\|_{\dot{H}^s\left(\mathbb{R}^d\right)}^2-\mathcal{C}_{s, d} \int_{\mathbb{R}^d} \frac{|\chi u|^2}{|x|^{2 s}} d x\right) \geq \xi^{-1}\|\chi u\|_{\dot{H}^t\left(\mathbb{R}^d\right)}^2 - \frac{C_{d,s,t}}{\xi^\frac{s+t}{s-t}}\|\chi u\|_{L^2\left(\mathbb{R}^d\right)}^2,
	\end{equation*} 
	where $t$  will be chosen later. Thus,  we combine   the former inequality with 	\eqref{HGN-inequality} that   for all $u \in H^s\left(\mathbb{R}^d\right)$,  we get 
	\begin{align}\label{esti-chi u-C-HGN-u-rest}
		&\|\chi u\|_{\dot{H}^s\left(\mathbb{R}^d\right)}^2-\mathcal{C}_{s, d} \int_{\mathbb{R}^d} \frac{|\chi u|^2}{|x|^{2 s}} d x  \nonumber\\
		&\geq C_{\mathrm{HGN}}(1-\xi) \frac{\int_{\mathbb{R}^d}|\chi u|^{2(1+2 s / d)}}{\left(\int_{\mathbb{R}^d}|\chi u|^2\right)^{2 s / d}}+\xi^{-1}\|\chi u\|_{\dot{H}^t\left(\mathbb{R}^d\right)}^2  -\frac{C_{d,s,t}}{\xi^\frac{s+t}{s-t}}\|\chi u\|_{L^2\left(\mathbb{R}^d\right)}^2 \nonumber\\ 
		&\geq  C_{\mathrm{HGN}}(1-\xi) \frac{\int_{\Omega}| u|^{2(1+2 s / d)}}{\left(\int_{\tilde \Omega}| u|^2\right)^{2 s / d}}+\xi^{-1}\|\chi u\|_{\dot{H}^t\left(\mathbb{R}^d\right)}^2 -\frac{C_{d,s,t}}{\xi^\frac{s+t}{s-t}}\| u\|_{L^2\left(\mathbb{R}^d\right)}^2 	\end{align}
	since ${\rm supp}(\chi) \subset \tilde \Omega$ and $\chi =1$ on $\Omega$.  Now, we claim that for all $ 0 < s < \frac{d}{2}$, 
	\begin{equation}\label{esti-chi-u-le-chi-u-H-s-dot-dot-H-t-1}		\|\chi u\|_{\dot{H}^s(\widetilde{\Omega})}^2 \leq\|\chi u\|_{\dot{H}^s\left(\mathbb{R}^d\right)}^2 \leq\|\chi u\|_{\dot{H}^s(\widetilde{\Omega})}^2 + \frac{1}{2}\|\chi u\|_{\dot{H}^{ t_1}\left(\mathbb{R}^d\right)}^2 + C(d,s) \|u\|^2_{L^2(\tilde \Omega)}
	\end{equation} 
	and 
	\begin{equation}\label{esti-chi-u-le-chi-u-H-s-dot-dot-H-t-2}
		\|\chi u\|_{\dot{H}^s(\widetilde{\Omega})}^2 \leq\|u\|_{\dot{H}^s(\widetilde{\Omega})}^2 + \frac{1}{2}\|\chi u\|_{\dot{H}^{t_1}\left(\mathbb{R}^d\right)}^2 +C(d,s)(1 + |\tilde \Omega|^\frac{2\sigma}{d})^{\frac{t_1}{t_1 -t_0}}\|u\|_{L^2(\widetilde{\Omega})}^2.
	\end{equation}
		
	\smallskip 
	\noindent
	{\bf Proof of \eqref{esti-chi-u-le-chi-u-H-s-dot-dot-H-t-1}:}   If $s \in \N $, then  \eqref{esti-chi-u-le-chi-u-H-s-dot-dot-H-t-1} holds true without the  presence of the second term. Let $s \notin \N$, i.e. $\sigma(s) \in  (0,1)$, we have the obvious bound  
	$$ \|  \chi u \|^2_{\dot H^s(\tilde \Omega)}   \le \|  \chi u\|^2_{\dot H^s(\R^d)}.    $$ 
	Then using the fact that $m \le t_1$,   \eqref{control-f-H-m-H-sL2} and  \eqref{compa-chi-u-R-d-chi-u-tilde-Omega-u-H-m},    we arrive at   \eqref{esti-chi-u-le-chi-u-H-s-dot-dot-H-t-1} since
	\begin{align*}
		\| \chi u\|^2_{\dot H^s(\R^d)} & \le \| \chi u\|^2_{\dot H^s(\tilde \Omega)}  + C(d,s) \| \chi u\|^2_{\dot H^m(\R^d)}\\
		&\le \| \chi u\|^2_{\dot H^s(\tilde \Omega)}  +  \frac{1}{2}\| \chi u\|^2_{\dot H^{t_1}(\R^d)} + C(d,s) \|u\|^2_{L^2(\tilde \Omega)}. 
	\end{align*}
	
	\smallskip 
		\noindent
		{\bf Proof of \eqref{esti-chi-u-le-chi-u-H-s-dot-dot-H-t-2}:}  We can rewrite \eqref{norm-u-chi-eta-u-le-t-0-notinteger}  as 
	\begin{align*}
		\left|\|u\|_{\dot{H}^s(\tilde \Omega)}^2-\|\chi u\|_{\dot{H}^s(\tilde \Omega)}^2-\|\eta u\|_{\dot{H}^s(\tilde \Omega)}^2\right| 
		\le C(d,s)(1 +|\tilde \Omega|^{\frac{2\sigma}{d}})\left(\|\chi u\|_{H^{t_0}(\tilde \Omega)}^2+\|\eta u\|_{H^{t_0}(\tilde \Omega)}^2\right). %prove-Hardy
	\end{align*} 
Applying  	
	\eqref{compa-H-t-dot-H-s-impro} we find that 
	\begin{align*}
		C(d,s) (1 +|\tilde \Omega|^{\frac{2\sigma}{d}})\| \eta u\|^2_{H^{t_0}(\tilde \Omega)} &\le \| \eta u \|^2_{\dot H^s(\tilde \Omega)} + C_1(d,s)(1 + |\tilde \Omega|^\frac{2\sigma}{d})^{\frac{s}{s-t_0}}\| \eta u \|^2_{L^2(\tilde \Omega)}, \\
		C(d,s) (1 +|\tilde \Omega|^{\frac{2\sigma}{d}})\| \chi u\|^2_{H^{t_0}(\tilde \Omega)} &\le \frac{1}{2} \| \chi u \|^2_{\dot H^{t_1}(\tilde \Omega)} + C_1(d,s)(1 + |\tilde \Omega|^\frac{2\sigma}{d})^{\frac{t_1}{t_1-t_0}}\| \eta u \|^2_{L^2(\tilde \Omega)}\\
		&\le \frac{1}{2} \| \chi u \|^2_{\dot H^{t_1}(\R^d)} + C_1(d,s)(1 + |\tilde \Omega|^\frac{2\sigma}{d})^{\frac{t_1}{t_1-t_0}}\| \chi u \|^2_{L^2(\tilde \Omega)}.
	\end{align*}
	So, by the triangle inequality, we get the conclusion of \eqref{esti-chi-u-le-chi-u-H-s-dot-dot-H-t-2}. 
	
	Finally, inserting  \eqref{esti-chi-u-le-chi-u-H-s-dot-dot-H-t-1} and \eqref{esti-chi-u-le-chi-u-H-s-dot-dot-H-t-2}  in  \eqref{esti-chi u-C-HGN-u-rest} we conclude that 
	\begin{align*}
			\| u\|_{\dot{H}^s\left(\mathbb{R}^d\right)}^2-\mathcal{C}_{s, d} \int_{\mathbb{R}^d} \frac{| u|^2}{|x|^{2 s}} d x  & \geq   C_{\mathrm{HGN}} (1-\xi)\frac{\int_{\Omega}| u|^{2(1+2 s / d)}}{\left(\int_{\tilde \Omega}| u|^2\right)^{2 s / d}} -\frac{C(d,s,t)}{\xi^\frac{s+t}{s-t}}\| u\|_{L^2\left(\tilde \Omega\right)}^2 \nonumber\\ 
		+ \xi^{-1}\|\chi u\|_{\dot{H}^t\left(\mathbb{R}^d\right)}^2  & -   \| \chi u\|^2_{\dot H^{t_1}(\R^d)}    -  C(d,s)(1+|\tilde \Omega|^\frac{2\sigma}{d})^\frac{t_1}{t_1-t_0}  \|u\|^2_{L^2(\tilde \Omega)}.\nonumber
	\end{align*}
	Choosing $t=t_1$, we have for all  $\xi \in (0,1)$,    
	$$\xi^{-1}\|\chi u\|_{\dot{H}^t\left(\mathbb{R}^d\right)}^2  -   \| \chi u\|^2_{\dot H^{t_1}(\R^d)} \ge 0.$$
	The desired inequality  \eqref{one-body-hardy-unvertainty} thus follows. The proof of Lemma \ref{lemma-for-Uncertainty} is complete.
\end{proof}

\section{Local exclusion estimates}\label{section-exclusion-principle}

In this section,  we  give some local versions of the exclusion principle associated with    the inverse nearest-neighbor interaction.  Our aim is to give a good lower bound for the interaction energy 
	\begin{equation}\label{defi-mathcal-I-R-d}
		\mathcal{I}_{\R^d}(\Psi)=\int_{\R^{dN}}  \sum_{i=1}^N \frac{|\Psi(x_1,...,x_N)|^2}{\delta^{2s}_i(x_1,...,x_N)} \prod_{j=1}^N \,dx_j .
	\end{equation}
	As in Section \ref{section-local-uncertainty-principles}, we always assume that $\Psi$ is a normalized wave function in $L^2(\R^{dN})$ such that $\Psi\in H^s(\R^{dN})$ and its one-body density $\rho_\Psi$ is supported in $[-1/2,1/2]^{d}$. We construct the covering sub-cubes of $[-1/2,1/2]^{d}$ as in Section \ref{section-subcovering}, using $\rho=\rho_\Psi$ and two parameters $\delta \in (0,1)$ and    
$\varepsilon \in  \{\frac{1}{2}, \frac{1}{3}\}$. 	Recall that for every cluster $K$ in $G^{n,1}$, we define  $\Omega_{K}$ as in \eqref{defi-Omga-K-tau}, and define the enlarged set $\tilde \Omega_{K}$ of $\Omega_{K}$ as in Lemma \ref{lem-tilde-Omega-K}. 
	
		The main result of this section is the following:  
	
	\begin{lemma}\label{lem:interaction} For all $d\in \mathbb{N}$ and $s>0$, there exists a universal constant $C=C(d,s)>0$ such that 
		\begin{equation}\label{est-mathcal-I-R-d}
	\mathcal{I}_{\R^d}  (\Psi) \ge 	\sum_{n \ge 1}   \frac{\delta^{1+2s}}{C\varepsilon^{2sn}} \left( \sum_{Q \in G^{n,0}} \int_Q \rho_\Psi + \sum_{K \subset G^{n,1}} \int_{\tilde \Omega_K}  \rho_\Psi \right). 
\end{equation}
	\end{lemma}

Let us start by proving a preliminary lower bound, using the homogeneity of the inverse nearest-neighbor interaction. This is based on an idea from  \cite{KogNam-21} and allows us to combine the interaction energy in different length scales. 

\begin{lemma}\label{lema-Omega-nm-int-Omega-local-exclusion-II}
	Let  $\{ R_n \}_{n \ge 1}$ be a decreasing sequence of positive numbers, $\{ C_n \}_{n \ge 1}$ a sequence of positive numbers,  and $ \{ \Omega_{n,m}\}$ a collection of subsets  of $\R^d$ satisfying 
	\begin{equation}\label{condition-diam-Omega-nm-sum-1-Omega-le-C-nm}
		{\rm diam}(\Omega_{n,m})  \le R_n \quad \text{ for all } m,  \quad \sum_{m \ge 1} \1_{\Omega_{n,m}}  \le  C_n \quad  \text{ for all } n. 
	\end{equation}
	Then for every normalized wave function $\Psi \in L^2(\R^{dN})$ we have
	\begin{equation}
		\mathcal{I}_{\R^d}  (\Psi)  \ge \sum_{n \ge 1} \frac{1}{2C_n} \left( \frac{1}{R^{2s}_n} - \frac{1}{R^{2s}_{n-1}} \right)  \sum_{m \ge 1} \left( \int_{\Omega_{n,m}} \rho_\Psi -1 \right).
	\end{equation}   
	Here we denote   $R_0 = +\infty$  for convenience. 
\end{lemma} 
\begin{proof} By the definition of $\delta_i= \delta_i(x_1,...,x_N)$  in \eqref{eq:W-nearest}, we have for all $i \in \{1,...,N\}$,
	\begin{align*}
		\frac{1}{\delta_i^{2s}} \ge \sum_{n \ge 1} \frac{\1_{\{ R_{n+1} < \delta_i \le R_n \}}}{\delta_i^{2s}} 
		\ge  \sum_{n \ge 1} \frac{\1_{ \{ \delta_i \le R_n \} } - \1_{ \{ \delta_i \le R_{n+1} \} } }{R_n^{2s}}  \ge \sum_{n \ge 1} \left( \frac{1}{R^{2s}_n} - \frac{1}{R^{2s}_{n-1}} \right) \1_{\{\delta_i \le R_n\}},  
	\end{align*} 
	with the convention  $R_0 = +\infty$. 	Moreover, we deduce from \eqref{condition-diam-Omega-nm-sum-1-Omega-le-C-nm}
	that for each $n \ge 1$, 
	\begin{align*}
		\sum_{i=1}^N \1_{ \{ \delta_i \le R_n \} } \ge  \frac{1}{C_n} \sum_{m \ge 1} \sum_{i=1}^N  \1_{ \{ \delta_i \le R_n \} } \1_{\Omega_{n,m}}(x_i) \ge  \frac{1}{C_n} \sum_{m \ge 1} \left(\sum_{i=1}^N \1_{\Omega_{n,m}}(x_i)- 1  \right). 
	\end{align*}
	Thus,  	\begin{align*}
		\mathcal{I}_{\R^d}(\Psi) &\ge \sum_{n \ge 1} \frac{1}{C_n} \left( \frac{1}{R_n^{2s}} - \frac{1}{R^{2s}_{n-1}} \right)  \sum_{m \ge 1} \left\langle  \Psi,  \left(\sum_{i=1}^N \1_{\Omega_{n,m}}(x_i)-1 \right) \Psi  \right\rangle \\
		& = \sum_{n\ge 1}\frac{1}{C_n} \left( \frac{1}{R_n^{2s}} - \frac{1}{R^{2s}_{n-1}} \right)  \sum_{m \ge 1} \left( \int_{\Omega_{n,m}} \rho_{\Psi} -1 \right),
	\end{align*}
	which concludes the proof of Lemma \ref{lema-Omega-nm-int-Omega-local-exclusion-II}. 
\end{proof}

Now let us consider a sub-cube  $Q\in G^{n,2}$, for some $n \ge 1$, and denote by  $c_Q$ the centered of $Q$. Since $Q \in G^{n,2}$, it  belongs to  a cluster $K  \subset G^{n,2}$ and we have
\begin{align*}
	\int_{\tilde \Omega_K} \rho_{\Psi} \ge 1 + \delta,\quad \text{ and } \int_{Q'} \rho_{\Psi} \ge \delta \quad \text{ for all } Q' \in K. 
\end{align*} 
Since $K$ contains connected sub-cubes of diameter $\sqrt{d} \varepsilon^n$ and $\tilde \Omega_K \subset \Omega_{K,\frac 1 2}$, if we choose 
\begin{align}\label{eq:def-Rn}
R_n = 8 \sqrt{d}(\delta^{-1} + 3) \varepsilon^n,
\end{align}
then the ball $B\left(c_Q, \frac{R_n}{4}\right)$ contains either $\tilde \Omega_K$,  or at least $[\delta^{-1}] +2$ disjoint sub-cubes  of $K$. In both cases, we always have  
\begin{equation}\label{integral-rho-B-R-n-2-greater-than-1+delta}
	\int_{B\left(c_Q, \frac{R_n}{4}\right)} \rho_\Psi \ge 1+  \delta \text{ for all } Q \in G^{n,2}.
\end{equation}  

%Indeed, let  $Q \in G^{n,2}$, then it  belongs to     a cluster $K  \subset G^{n,2}$. Besides that, the diameter of each  $Q\in G^{n,2}$ has length $\sqrt{d} \varepsilon^n$. So, either $\Omega_K \subset B\left(c_Q, \frac{R_n}{2}\right)$  or the ball contains at least $[\delta^{-1} +2]$ disjoint sub-cubes  of $K$.  Moreover,    the construction  of  $G^{n,2}$ implies that 
%\begin{align*}
%	\int_{\Omega_K} \rho_{\Psi} \ge 1 + \delta \text{ and } \int_{Q} \rho_{\Psi} \ge \delta \text{ for all } Q \in G^{n,2} \text{ and  cluster } K \subset G^{n,2}. 
%\end{align*}
%Thus, \eqref{integral-rho-B-R-n-2-greater-than-1+delta} follows. 

Next, we want to cover the set  
$$ E := \bigcup_{Q \in G^{n,2}} Q_{\frac 1 2},$$
by a collection of balls, where $Q_{\tau}$ is defined in \eqref{defi-Omga-K-tau}. Obviously, we can cover $E$ by the balls $\left\{B\left(c_Q, \frac{R_n}{4}\right)\right\}_{Q \in G^{n,2}}$. However, these balls may overlap too much, and we will avoid this by using the Besicovitch covering lemma. 

\begin{lemma}[Besicovitch covering lemma  \cite{Besicovitch-45}]\label{Besicovitch-lemma}
	Let $E$ be a bounded subset of $\mathbb{R}^d$ and  $\mathcal{F}$ be a collection of balls in $\mathbb{R}^d$ such that every point $x \in E$ is contained in at least   a ball $B$ from $\mathcal{F}$. Then, we can find  a sub-collection $\mathcal{G} \subset \mathcal{F}$ such that 
	\[
	\1_E \leq \sum_{B \in \mathcal{G}} \1_B \leq C(d),
	\]
	where the constant $C(d) > 0$ depends only on the dimension $d \geq 1$. % (in particular, it  is independent of $E$).
\end{lemma}

\smallskip 
\noindent 
%exclusion-nearest-neighbor

Now we are ready to conclude Lemma \ref{lem:interaction}. 

\begin{proof}[Proof of Lemma \ref{lem:interaction}]
We observe that for each point $x \in E =\bigcup_{Q \in G^{n,2}} Q_{\frac 1 2}$, there exists a sub-cube $Q \in G^{n,2}$ such that $x \in Q_{\frac 1 2}$. By the triangle inequality,
$$   B\left(c_Q, \frac{R_n}{4}\right)   \subset  B\left(x, \frac{R_n}{2}\right).$$
Here recall that $Q$ is centered at $c_d$, ${\rm diam}(Q)=\sqrt{d} \varepsilon^n$, $Q_{\tau}$ is defined in \eqref{defi-Omga-K-tau}, and $R_n$ is chosen in \eqref{eq:def-Rn}. Therefore, by \eqref{integral-rho-B-R-n-2-greater-than-1+delta}, 
\begin{equation}\label{integral-rho-B-R-n-2-greater-than-1+delta-3-2}
	\int_{B\left(x, \frac{R_n}{2}\right)} \rho_\Psi \ge 1+  \delta, \quad \text{ for all } x \in E.
\end{equation} 

Applying Lemma \ref{Besicovitch-lemma} to $\mathcal{F}=\Big\{B\left(x, \frac{R_n}{2}\right)\Big\}_{x\in E}$, we can find a collection of the balls  $B_{n,m} = B\left(x_{n,m},\frac{R_n}{2}\right)$, with  $\{x_{n,m}\}_{m \ge 1} \subset E$, such that   
$$  \1_E \le	\sum_{m \ge 1} \1_{B_{n,m}} \le  C(d).$$
%Let us define the family  
%$$  \left\{ B\left( x, \frac{3}{2}R_n \right) , x \in E\right\}.$$
%
%
%
%\smallskip
%\noindent 
Now, we apply Lemma  \ref{lema-Omega-nm-int-Omega-local-exclusion-II} with $R_n$ given in \eqref{eq:def-Rn}, $\Omega_{n,m}= B_{n,m}$ and $C_n=C(d)$ (independent of $n$). We find that 
%with 
%$$ R_n = {\rm diam}\left(B\left({c_Q,\frac{R_n}{2}}\right) \right)  = 2 \sqrt d \left(\delta^{-1} + 3 \right) \varepsilon^n,  $$
%that it follows 
\begin{align}\label{esti-exclusion-on-G-n-2}
	\mathcal{I}_{\R^d}(\Psi)  &\ge   \sum_{n \ge 1} \frac{1}{2C(d)} \left( \frac{1}{R^{2s}_n} - \frac{1}{R^{2s}_{n-1}} \right)  \sum_{m \ge 1} \left( \int_{B_{n,m}} \rho_\Psi -1 \right) \nonumber\\
	& \ge\sum_{n \ge 1} \frac{\delta }{C(d,s) (\delta^{-1} +3)^{2s}\varepsilon^{2sn}} \sum_{m\ge 1} \int_{B_{m,n}}\rho_\Psi.
	\end{align}
Here in the last estimate, we have used 
$$
\int_{B_{n,m}}\rho_\Psi -1 \ge \frac{\delta}{1+\delta} \int_{B_{n,m}}\rho_\Psi,
$$
which is equivalent to $\int_{B_{n,m}}\rho_\Psi \ge 1 +\delta$, a consequence of \eqref{integral-rho-B-R-n-2-greater-than-1+delta-3-2}. To go further, we use the fact that the balls $\{B_{n,m}\}_{m\ge 1}$ cover $E=\bigcup_{Q \in G^{n,2}} Q_{\frac 1 2}$, which implies that 
\begin{align*}
	\left(\bigcup_{Q \in Q^{n+1,0} } Q \right)  \bigcup  \left( \bigcup_{ K \subset  G^{n+1,1} }  \Omega_{K,\frac 1 2} \right)  \subset  	\bigcup_{m \ge 1} B_{n,m}. 
\end{align*}
On the left-hand side, note that $ \tilde \Omega_K\subset \Omega_{K,\frac 12}$. Since the sub-cubes $Q $ in $G^{n+1,0}$ are disjoint and the sets $\tilde \Omega_K$ with different clusters $K$ in $G^{n+1,1}$ are also disjoint, we deduce that 
\begin{align*}
	\sum_{m\ge 1}  \int_{B_{n,m}} \rho_\Psi \ge \frac 1 2 \left( \sum_{Q \in G^{n+1,0}} \int_{Q} \rho_{\Psi} + \sum_{K \subset G^{n+1,1}} \int_{\tilde \Omega_K} \rho_{\Psi}   \right).
\end{align*}
Inserting the latter bound in \eqref{esti-exclusion-on-G-n-2}, we obtain the desired conclusion of Lemma \ref{lem:interaction}. 
\end{proof}

\section{Lieb--Thirring inequalities}\label{section-proo-theorem-1}

Now we are ready to prove Theorem \ref{thm:1-fractional}. 
%The overall strategy is similar to that of   \cite{KogNam-21}, but we will keep track all quantitative error estimates. In particular, to combine the local uncertainty and exclusion estimates in an efficient way, we will make use the Besicovitch covering lemma.  

\smallskip 
\noindent 
\begin{proof}[Proof of Theorem \ref{thm:1-fractional}] Let $d\in \mathbb{N}$ and $s>0$. First, the upper bound on $K_{\rm LT}(d,s,\lambda)$ can be obtained by considering the trial state 
\begin{align}\label{eq:upper-bound-trial-state}
\Psi_\ell (x_1,x_2,...,x_N)= \prod_{j=1}^N u(x_j - \ell j x_0)
\end{align}
where $x_0$ is a unit vector in $\R^d$, $u$ is a normalized function in $L^2(\R^2)$ that is smooth and compactly supported, and $\ell>0$ is a large parameter. When $\ell\to \infty$, the wave function $\Psi_\ell$ describes a state of $N$ particles, each very far from the others. Consequently,
\begin{align}\label{eq:upper-bound-trial-state-2}
 \lim_{\ell \to \infty} \frac{ \left\langle \Psi, \left( \sum_{i=1}^N (-\Delta_{x_i})^s_{| \R^d} \right) \Psi\right\rangle  +  \lambda \mathcal{I}_{\R^d}(\Psi)}{\int_{\R^d} \rho_\Psi^{1+2s/d}} = \dfrac{\langle u, (-\Delta )^s u\rangle}{\int_{\R^d} |u|^{2(1+{\frac {2s} d})}}.
\end{align}
Note that \eqref{eq:upper-bound-trial-state-2} is valid for every fixed $N\ge 1$. Optimizing over $\|u\|_{L^2}=1$ we obtain $K_{\rm LT}(d,s,\lambda)\le C_{\rm GN}(d,s)$.

\bigskip

To prove \eqref{eq:LT-nearest-potential-1-fractional} and \eqref{eq:LT-nearest-potential-2-fr}, by a density argument, it suffices to consider an arbitrarily normalized wave function $\Psi\in  L^2(\mathbb{R}^{dN})$ which is smooth and compactly supported. Moreover, by a scaling argument, we may assume that  \( \Psi \) is supported in \( \left[-\frac{1}{2}, \frac{1}{2}\right]^{Nd} \). Consequently, its density $\rho_\Psi$ is supported in $ \left[-\frac{1}{2}, \frac{1}{2}\right]^d$. Then we construct the covering sub-cubes as in Section \ref{section-subcovering} with $\rho=\rho_\Psi$, $\delta\in (0,1)$ and $\eps=\frac 1 2$ (or $\eps=\frac 1 3$). In particular, we have
\begin{equation}\label{eq:suppor-rho-psi-Gn}
	\text{supp}(\rho_\Psi) \subset \bigcup_{n \ge 1 } \left( \bigcup_{Q \in G^{n,0}\cup G^{n,1}} Q \right). 
\end{equation}

In the following, we will collect separately the kinetic energy estimate for sub-cubes in $G^{n,0}$ and the kinetic energy estimate for sub-cube in $G^{n,1}$. Then we combine them with the interaction energy estimate to get the final conclusion.

\bigskip 
\noindent
{\bf Uncertainty principle  for $G^{n,0}$.}  Recall that for all $n \ge 1$, if   $Q \in G^{n,0}$, then  
$$ \int_Q \rho_{\Psi}  \le \delta \quad \text{ and } \quad |Q| = \varepsilon^{nd}.$$%uncertainty-G-n-0
Hence, we deduce from Lemma \ref{lemma-local-uncertainty-II}, Eq. \eqref{uncertainty-type-I},  that 
\begin{align*}
	\left\langle \Psi, \left( \sum_{i=1}^N (-\Delta_{x_i})^s_{| Q} \right) \Psi\right\rangle 
	\ge \frac{1}{C\delta^{\frac{2s}{d}}} \int_Q \rho_{\Psi}^{1 + \frac{2s}{d}} - \frac{C}{\varepsilon^{2sn}} \int_Q \rho_{\Psi},
\end{align*}
with a universal constant $C = C(d,s)>0$ depending only on \( d \) and \( s \) (this notation will be used throughout the rest of the paper). Moreover, since the subcubes in $G^{n,0}$ are disjoint, we have   
\begin{align}\label{uncertainty-principle-G-n-0}
	C\delta^{2s/d} \left\langle\Psi, \sum_{i=1}^N\left(-\Delta_{x_i}\right)_{\mid \mathbb{R}^d}^s \Psi\right\rangle  \geq \sum_{n \geq 1} \sum_{Q \in G^{n, 0}}\left( C_{\rm GN} \int_Q \rho_{\Psi}^{1+ \frac{2s}{d}}-\frac{C  \delta^\frac{2s}{d} }{\varepsilon^{2 s n}} \int_Q \rho_{\Psi}\right). 
\end{align}
for a universal constant $C=C(d,s)>0$. 
%lower bound 
%We will use the lower bound in \eqref{uncertainty-principle-G-n-0} to control the localization error, and hence it is fine to obtain a non-sharp constant factor. 
 
\bigskip 
\noindent 
{\bf Uncertainty principle for $G^{n,1}$.} %uncer-G-n-1-Laplace
Let $K$ be a cluster in $G^{n,1}$. Then  
\begin{equation}\label{eq:cluster-K-Q-1}
	\int_Q \rho_{\Psi}  \ge \delta, \text{  for all } Q \in K,
\end{equation}
and 
\begin{equation}\label{eq:cluster-K-Q-2}
	\sum_{Q \in K} \int_Q \rho_{\Psi} = \int_{\Omega_K} \rho_{\Psi} \le \int_{\tilde \Omega_K} \rho_\Psi < 1 + \delta, 
\end{equation}
where \( \Omega_K \) is defined in \eqref{defi-Omga-K-tau} and \( \tilde{\Omega}_K \) is determined in Lemma \ref{lem-tilde-Omega-K}. From \eqref{eq:cluster-K-Q-1} and \eqref{eq:cluster-K-Q-2}, we deduce that $|\# K| \le [\delta^{-1} ] +2$, namely \( K \) contains at most \( ([\delta^{-1}] + 2) \) disjoint sub cubes. Thus applying  Lemma \ref{lemma-local-uncertainty-II}, Eq. \eqref{esti-local-uncertainty-II}, with $\xi=\delta$, we find that 
% In addition, we observe that 
%\begin{equation}\label{defi-varep-n-Omega-K}
%	\varepsilon^{-n} \Omega_{K}  = \bigcup_{Q \in K} (\varepsilon^{-n}Q)= \Omega_{K_0},
%\end{equation} 
%where $K_0$ is a cluster of connected unit cubes.  Thus, we apply \eqref{esti-local-uncertainty-II} for $\Psi_\eps$ defined by 
%\begin{align*}
%	\Psi_{\varepsilon}(x)  = \varepsilon^{\frac{dn}{2}} \Psi \left( x\varepsilon^n \right),	
%\end{align*}	 
%that we derive 
%\begin{equation*}
%	\left\langle \Psi, \left( \sum_{i=1}^N (-\Delta_{x_i})^s_{| \tilde \Omega_K} \right) \Psi\right\rangle \ge  C_{ \rm GN} (1 - \delta)\frac{\int_{\Omega_K}  \rho_{\Psi}^{1+\frac{2s}{d}}}{\left(\int_{\tilde \Omega_K}  \rho_{\Psi} \right)^\frac{2s}{d}} - \frac{C(1+ |\# K_0|^{\frac{2\sigma}{d}})^{\frac{s}{s-t_0}}}{\epsilon^{2sn}\delta^{\frac{t_0}{s-t_0}} }\int_{ \tilde \Omega_{K_0}} \rho_{\Psi}.
%\end{equation*}%G-n-1
%Note that $|\# K| \le [\delta^{-1} ] +2$ and $\delta \in (0,1)$,   we obtain 
\begin{align}
	\left\langle \Psi, \left( \sum_{i=1}^N (-\Delta_{x_i})^s_{| \tilde \Omega_K} \right) \Psi\right\rangle \ge  C_{ \rm GN} (1 - \delta)\frac{\int_{\Omega_K}  \rho_{\Psi}^{1+\frac{2s}{d}}}{\left(\int_{\tilde \Omega_K}  \rho_{\Psi} \right)^\frac{2s}{d}} - \frac{C}{\epsilon^{2sn}\delta^{\eta_1} }\int_{ \tilde \Omega_{K}} \rho_{\Psi}.\label{esti-mathcal-T-on-G-n-0and-1-1}
\end{align}
with 
\begin{equation}\label{defi-eta-1}
	\eta_1 =  \frac{2\sigma(s)s}{d(s-t_0)} + \frac{t_0}{s-t_0} \ge 0
\end{equation} 
($t_0$ is given in \eqref{defi-t-0}). Moreover, since  $\tilde \Omega_{K_1}$ and $\tilde \Omega_{K_2}$  are  disjoint  for two different clusters $ K_1, K_2$ in $G^{n,1}$, we can sum \eqref{esti-mathcal-T-on-G-n-0and-1-1} over all clusters $K$ in $G^{n,1}$ and arrive at 
\begin{align}\label{esti-mathcal-T-on-G-n-0and-1}
	&	\left\langle \Psi, \left( \sum_{i=1}^N (-\Delta_{x_i})^s_{| \R^d} \right) \Psi\right\rangle \ge  \sum_{n \ge 1} \sum_{K \subset G^{n,1}} \left\langle \Psi, \left( \sum_{i=1}^N (-\Delta_{x_i})^s_{| \tilde \Omega_K} \right) \Psi\right\rangle \nonumber \\
	&	\ge \sum_{n \ge 1} \sum_{K \subset G^{n,1} }  \left(C_{\rm GN}(1 -\delta) \frac{\int_{\Omega_K} \rho_\Psi^{1 + \frac{2s}{d}}}{\left( \int_{\tilde \Omega_K} \rho_{\Psi} \right)^\frac{2s}{d}} - \frac{C}{\varepsilon^{2sn} \delta^{\eta_1} } \int_{ \tilde \Omega_{K}} \rho_{\Psi} \right) \nonumber\\
	& \ge  \sum_{n \ge 1} \sum_{K \subset G^{n,1} }  \left(\frac{C_{\rm GN}(1 -\delta)}{(1 +\delta)^\frac{2s}{d}}  \int_{\Omega_K} \rho_\Psi^{1 + \frac{2s}{d}} - \frac{C}{\varepsilon^{2ns} \delta^{\eta_1}} \int_{ \tilde \Omega_{K}} \rho_{\Psi} \right).
\end{align} 
Here we have used \eqref{eq:cluster-K-Q-2} in the last estimate. 

%Consequently,
%\begin{align}\label{esti-mathcal-T-on-G-n-0and-1}
%	& (1-C \delta^{2s/d})	\left\langle \Psi, \left( \sum_{i=1}^N (-\Delta_{x_i})^s_{| \R^d} \right) \Psi\right\rangle \ge  \sum_{n \ge 1} \sum_{K \subset G^{n,1}} \left\langle \Psi, \left( \sum_{i=1}^N (-\Delta_{x_i})^s_{| \tilde \Omega_K} \right) \Psi\right\rangle \nonumber \\
%	&\ge \sum_{n \ge 1} \sum_{K \subset G^{n,1} }  \left(C_{\rm GN}(1 -\delta) \frac{\int_{\Omega_K} \rho_\Psi^{1 + \frac{2s}{d}}}{\left( \int_{\tilde \Omega_K} \rho_{\Psi} \right)^\frac{2s}{d}} - \frac{C}{\varepsilon^{2sn} \delta^{\eta_1} } \int_{ \tilde \Omega_{K}} \rho_{\Psi} \right) \nonumber\\
%	& \ge  \sum_{n \ge 1} \sum_{K \subset G^{n,1} }  \left(\frac{C_{\rm GN}(1 -\delta)}{(1 +\delta)^\frac{2s}{d}}  \int_{\Omega_K} \rho_\Psi^{1 + \frac{2s}{d}} - \frac{C}{\varepsilon^{2ns} \delta^{\eta_1}} \int_{ \tilde \Omega_{K}} \rho_{\Psi} \right).
%\end{align} 

%The right-hand side of \eqref{esti-mathcal-T-on-G-n-0and-1} is the main energy contribution, and hence it is important to keep the optimal constant $C_{\rm GN}$ here. 

\bigskip 
\noindent 
{\bf Conclusion.} Combining the uncertainty estimates \eqref{uncertainty-principle-G-n-0}, \eqref{esti-mathcal-T-on-G-n-0and-1} and the exclusion principle in Lemma \ref{lem:interaction}, we can find a universal constant $C=C(d,s)>0$ such that 
\begin{align*}
	&C\delta^{{\frac {2s} d}} \left\langle \Psi, \left( \sum_{i=1}^N (-\Delta_{x_i})^s_{| \R^d} \right) \Psi\right\rangle + (1-C\delta^{{\frac {2s} d}}) \left\langle \Psi, \left( \sum_{i=1}^N (-\Delta_{x_i})^s_{| \R^d} \right) \Psi\right\rangle +  \lambda \mathcal{I}_{\R^d}(\Psi)\\
	&\ge \sum_{n \geq 1} \sum_{Q \in G^{n, 0}}\left( C_{\rm GN} \int_Q \rho_{\Psi}^{1+ \frac{2s}{d}}-\frac{C  \delta^\frac{2s}{d} }{\varepsilon^{2 s n}} \int_Q \rho_{\Psi}\right) \\
	&\quad + (1-C\delta^{{\frac {2s} d}}) \sum_{n \ge 1} \sum_{K \subset G^{n,1} }  \left(\frac{C_{\rm GN}(1 -\delta)}{(1 +\delta)^\frac{2s}{d}}  \int_{\Omega_K} \rho_\Psi^{1 + \frac{2s}{d}} - \frac{C}{\varepsilon^{2ns} \delta^{\eta_1}} \int_{ \tilde \Omega_{K}} \rho_{\Psi} \right)\\
	&\quad + \lambda \sum_{n \ge 1}   \frac{\delta^{1+2s}}{C\varepsilon^{2sn}} \left( \sum_{Q \in G^{n,0}} \int_Q \rho_\Psi + \sum_{K \subset G^{n,1}} \int_{\tilde \Omega_K}  \rho_\Psi \right) \\
	&\ge  \sum_{n \geq 1} \left( \sum_{Q \in G^{n, 0}}  \int_Q \rho_{\Psi}^{1+ \frac{2s}{d}} + (1-C\delta^{{\frac {2s} d}}) \frac{C_{\rm GN}(1 -\delta)}{(1 +\delta)^\frac{2s}{d}}  \sum_{K \subset G^{n,1} } \int_{\Omega_K} \rho_{\Psi}^{1+ \frac{2s}{d}} \right) 
		\end{align*}
provided that 
$$
C\delta^{{\frac {2s} d}}<1,\quad  \lambda \delta^{1+2s} \ge \max \{ C \delta^{{\frac {2s} d}}, C \delta^{-\eta_1}\}. 
$$  
Choosing 
$$
\delta = \min \left\{ \frac 1 2 , \left( \frac 1 C \right)^{\frac d {2s}}, \left( \frac{C}{\lambda}\right)^{\frac 1 {1+2s+\eta_1}} \right\},
$$
then using $(1-C\delta^{{\frac {2s} d}})(1 -\delta) (1 +\delta)^{-\frac{2s}{d}} \ge 1- C(d,s) \delta^{\min \{1,2s/d\}}$ and \eqref{eq:suppor-rho-psi-Gn}, we conclude that 
\begin{align*}
	& \left\langle \Psi, \left( \sum_{i=1}^N (-\Delta_{x_i})^s_{| \R^d} \right) \Psi\right\rangle  +  \lambda \mathcal{I}_{\R^d}(\Psi)\\
	&\ge \left( C_{\rm GN} - C(d,s) \delta^{\min \{1,2s/d\}} \right)  \sum_{n \geq 1} \left( \sum_{Q \in G^{n, 0}}  \int_Q \rho_{\Psi}^{1+ \frac{2s}{d}} +  \sum_{K \subset G^{n,1} } \int_{\Omega_K} \rho_{\Psi}^{1+ \frac{2s}{d}} \right) \\
	&\ge \left( C_{\rm GN} - C(d,s) \lambda^{-k_1} \right) \int_{\R^d} \rho_{\Psi}^{1+ \frac{2s}{d}}, 
\end{align*}
provided that $\lambda$ is bigger than a universal constant $\lambda (d,s)>0$ and 
$$
k_1 = \frac{\min\{1,2s/d\}}{1+2s+\eta_1}
$$
with $\eta_1$ given in \eqref{defi-eta-1}. If $s=1$, then $\eta_1=0$, and hence $k_1 = \min\{\frac 1 3,  \frac 2 {3d}\}$. 

Thus we have proved \eqref{eq:LT-nearest-potential-1-fractional} and \eqref{eq:LT-nearest-potential-2-fr} in the case $\lambda>\lambda(d,s)$. In the case $\lambda \le \lambda_{d,s}$, we only need to prove  \eqref{eq:LT-nearest-potential-1-fractional}  for a constant $K_{\rm LT}(d,s,\lambda)\ge \frac{\lambda}{C_{d,s}}$, which can be deduce from the previous case by the obvious bound 
\begin{align}\label{eq:LH-small-lambda-trick}
	& \left\langle \Psi, \left( \sum_{i=1}^N (-\Delta_{x_i})^s_{| \R^d} \right) \Psi\right\rangle +  \lambda \mathcal{I}_{\R^d}(\Psi)\nonumber\\
	&\ge \frac{\lambda}{\lambda_{d,s}} \left( \left\langle \Psi, \left( \sum_{i=1}^N (-\Delta_{x_i})^s_{| \R^d} \right) \Psi\right\rangle +  \lambda(d,s) \mathcal{I}_{\R^d}(\Psi) \right).  
\end{align}
The proof of Theorem \ref{thm:1-fractional} is complete.
\end{proof}

\section{Hardy--Lieb--Thirring inequalities} \label{sec:HLT}\label{section-proof-theorem-2}

This section is devoted to the  proof of Theorem \ref{thm:2-fractional}. Since the strategy is the same to that of  Theorem \ref{thm:1-fractional}, we will only explain the key differences in handling the Hardy--Schrodinger operator. 

\begin{proof}[Proof of Theorem \ref{thm:2-fractional}] Let $d\in \mathbb{N}$ and $ 0  <  s < d/2$. The upper bound $K_{\rm HLT}(d,s,\lambda)\le C_{\rm HLT}(d,s)$ can be obtained by using the trial state in \eqref{eq:upper-bound-trial-state}. Therefore, we only focus on the lower bound on $K_{\rm HLT}(d,s,\lambda)$. Let  $\Psi$ be a normalized, smooth function supported in $\left[ -\frac{1}{2}, \frac{1}{2}\right]^{Nd}$. Then we use the construction in Section  \ref{section-subcovering} with $\rho=\rho_\Psi$, $\delta\in (0,1)$ and $\eps=\frac 1 3$. This choice of $\eps$ ensures that the 
covering sub-cubes  satisfy the following  additional property:  
\begin{equation}\label{proper-Q}
	\text{ for all } Q\in \bigcup_{n\ge 1}G^{n,0} \cup G^{n,1}, \text{  either $Q$ is centered at $0$, or  }    \text{ dist}\left(0,Q \right) \ge \frac{|Q|^\frac{1}{d}}{2}. 
\end{equation}
%uncertainty-G-n-0-hardy

\bigskip 
\noindent
First, we claim that for all $n \ge 1$ and  $Q \in G^{n,0} \cup G^{n,1}$, 
\begin{equation}\label{esti-Q-G-0-1-n-Har}
	\left\langle \Psi,   \sum_{i=1}^N \left( (-\Delta)^s_{|Q} - \frac{\mathcal{C}_{d,s}}{|x_i|^{2s}} \1_{Q}(x_i) \right) \Psi \right\rangle \ge  \frac{1}{C} \dfrac{\int_Q \rho_\Psi^{1+ \frac{2s}{d}}}{ (\int_Q \rho_\Psi )^{\frac{2s}{d}}} - \frac{C}{|Q|^\frac{2s}{d}} \int_Q \rho_\Psi.
\end{equation} 
Indeed, the case $0 \in Q$ has been covered by \eqref{uncertainty-I-hardy-type-ope}. If $0\notin Q$, then by \eqref{proper-Q} we have 
$$  \left\langle \Psi, \sum_{i=1}^N\frac{\mathcal{C}_{d,s}}{|x_i|^{2s}} \1_{Q}(x_i)  \Psi \right\rangle= \mathcal{C}_{d,s} \int_{Q} \frac{\rho_\Psi}{|x|^{2s}} dx \le  \frac{C}{|Q|^\frac{2s}{d}}  \int_{Q} \rho_\Psi,$$
and hence \eqref{esti-Q-G-0-1-n-Har} follows from \eqref{uncertainty-type-I}.  Moreover, since the   sub-cubes $Q \in \bigcup_{n\ge 1} G^{n,0} \cap G^{n,1}$ are disjoint and $\int_Q \rho_\Psi \le \delta $ provided that $Q \in G^{n,0}$, we have
\begin{align} \label{uncertainty-principle-G-n-0-hardy}
	&C\delta^\frac{2s}{d}  \left\langle \Psi,   \sum_{i=1}^N \left( (-\Delta)^s_{|\R^d} - \frac{\mathcal{C}_{d,s}}{|x_i|^{2s}} \right) \Psi \right\rangle  \nonumber \\
	&\geq \sum_{n \geq 1} \sum_{Q \in G^{n, 0} } C_{\rm HGN} \int_Q \rho_{\Psi}^{1+ \frac{2s}{d}}-  \sum_{n \geq 1} \sum_{Q \in G^{n, 0} \cup G^{n,1}} \frac{C \delta^\frac{2s}{d}  }{\varepsilon^{2 s n}} \int_Q \rho_{\Psi}.
\end{align}

\smallskip
Next, by Lemma \ref{lemma-for-Uncertainty} and the fact that $\int_{\tilde \Omega_K}\rho_\Psi<1+\delta$ for every cluster $K$ in $G^{n,1}$,  
\begin{align}\label{local-uncertainty-for-G-n-1-hardy}
	& \left\langle \Psi,   \sum_{i=1}^N \left( (-\Delta)^s_{|\R^d} - \frac{\mathcal{C}_{d,s}}{|x_i|^{2s}} \right) \Psi \right\rangle \nonumber\\
	&\ge \sum_{n \geq 1} \sum_{K \subset G^{n, 1}}\left\langle\Psi, \sum_{i=1}^N\left(\left(-\Delta_{x_i}\right)_{\mid \tilde{\Omega}_K}^s-\mathcal{C}_{s, d}\left|x_i\right|^{-2 s} \1_{\Omega_K}\left(x_i\right)\right) \Psi\right\rangle \nonumber\\ & \geq \sum_{n \geq 1} \sum_{K \subset G^{n, 1}}\left(C_{\mathrm{HGN}} \frac{(1-\delta)}{(1+\delta)^{\frac{2s}{d}}} \int_{\Omega_K} \rho_{\Psi}^{1+\frac{2s}{d}}-\frac{C}{\varepsilon^{2 s n} \delta^{\eta_2(s)}} \int_{\widetilde{\Omega}_K} \rho_{\Psi}\right),
\end{align}
where 
\begin{equation}\label{defi-eta_2}
	\eta_2(s) = \frac{s +t_1}{s-t_1} + \frac{2 \sigma(s) t_1}{d(t_1-t_0)}.
\end{equation} 
Recall that $t_0 $ and $t_1$  are defined  in  \eqref{defi-t-0} and  \eqref{defi-t-1}, respectively.

Combining \eqref{uncertainty-principle-G-n-0-hardy}, \eqref{local-uncertainty-for-G-n-1-hardy} and Lemma \ref{lem:interaction}, we have 
\begin{align*}
	&\left\langle \Psi,   \sum_{i=1}^N \left( (-\Delta)^s_{|\R^d} - \frac{\mathcal{C}_{d,s}}{|x_i|^{2s}} \right) \Psi \right\rangle   +  \lambda \mathcal{I}_{\R^d}(\Psi)\\
	&\ge \sum_{n \geq 1} \sum_{Q \in G^{n, 0}}\left( C_{\rm HGN} \int_Q \rho_{\Psi}^{1+ \frac{2s}{d}}-\frac{C  \delta^\frac{2s}{d} }{\varepsilon^{2 s n}} \int_Q \rho_{\Psi}\right) \\
	&\quad + (1-C\delta^{{\frac {2s} d}}) \sum_{n \ge 1} \sum_{K \subset G^{n,1} }  \left(\frac{C_{\rm HGN} (1 -\delta)}{(1 +\delta)^\frac{2s}{d}}  \int_{\Omega_K} \rho_\Psi^{1 + \frac{2s}{d}} - \frac{C}{\varepsilon^{2ns} \delta^{\eta_2}} \int_{ \tilde \Omega_{K}} \rho_{\Psi} \right)\\
	&\quad + \lambda \sum_{n \ge 1}   \frac{\delta^{1+2s}}{C\varepsilon^{2sn}} \left( \sum_{Q \in G^{n,0}} \int_Q \rho_\Psi + \sum_{K \subset G^{n,1}} \int_{\tilde \Omega_K}  \rho_\Psi \right) \\
	&\ge  \sum_{n \geq 1} \left( \sum_{Q \in G^{n, 0}}  \int_Q \rho_{\Psi}^{1+ \frac{2s}{d}} + (1-C\delta^{{\frac {2s} d}}) \frac{C_{\rm HGN} (1 -\delta)}{(1 +\delta)^\frac{2s}{d}}  \sum_{K \subset G^{n,1} } \int_{\Omega_K} \rho_{\Psi}^{1+ \frac{2s}{d}} \right) 
		\end{align*}
provided that 
$$
C\delta^{{\frac {2s} d}}<1,\quad  \lambda \delta^{1+2s} \ge \max \{ C \delta^{{\frac {2s} d}}, C \delta^{-\eta_2}\}. 
$$  
Choosing 
$$
\delta = \min \left\{ \frac 1 2 , \left( \frac 1 C \right)^{\frac d {2s}}, \left( \frac{C}{\lambda}\right)^{\frac 1 {1+2s+\eta_2}} \right\},
$$
and using \eqref{eq:suppor-rho-psi-Gn}, we conclude that when $\lambda$ is bigger than a universal constant $\lambda (d,s)>0$, 
\begin{align*}
	\left\langle \Psi, \left( \sum_{i=1}^N (-\Delta_{x_i})^s_{| \R^d} \right) \Psi\right\rangle  +  \lambda \mathcal{I}_{\R^d}(\Psi)
	\ge \left( C_{\rm HGN} - C(d,s) \lambda^{-k_2} \right) \int_{\R^d} \rho_{\Psi}^{1+ \frac{2s}{d}}, 
\end{align*}
where
$$
k_2= \frac{2s}{d(1+2s+\eta_2)}.
$$
Here note that we are considering $s<d/2$, and hence $\min\{1,2s/d\}=2s/d$. If $s=1$, then $t_0=t_1=0$, $\eta_2=1$ and $k_2=1/(2d)$. 

Thus we have proved \eqref{eq:HLT-nearest-potential-1-fractional} and \eqref{eq:LT-nearest-potential-2-fr} in the case $\lambda> \lambda(d,s)$. The case $\lambda\le \lambda(d,s)$ can be handled by the same argument in \eqref{eq:LH-small-lambda-trick}. The proof of  Theorem \ref{thm:2-fractional} is complete. 
\end{proof}

\appendix

\section{Fermionic estimates}\label{app}

In this section, we prove  \eqref{eq:theo-3} for all $0<s<d/2$ by following the method in \cite{LieYau-88}. First, we establish the following result. 

\begin{lemma}\label{lemma-Tr-gamma-Delta-s-V}  Let $d\in \mathbb{N}$ and $ 0<s < \frac{d}{2}$. Let $X_1,...,X_M$ be  $M$ distinct points in  $\R^d$. Then for all $0<\beta<\mathcal{C}_{d,s}$, with  $\mathcal{C}_{d,s}$ the constant in \eqref{Frac-Hardy-constant},  and for all density matrix $0 \le \gamma \le 1$, we have
	\begin{equation}\label{inequality-Tr-gamma-ge-alpha-R-j}
		\operatorname{Tr}\left(\left((-\Delta)^s-V\right)\gamma\right) \geq- C(d,s) \beta^{1+\frac d {2s}} \sum_{j=1}^M\left(2 R_j\right)^{-2 s}
	\end{equation}
	where $V:\R^d \to \R$ and $R_j\in (0,\infty)$  are defined by
	\begin{align*}
		V(x) & = \delta^{-2 s}(x),  \text{ with } \delta(x)  =\min \left\{\left|x-X_j\right|, j=1, \ldots, M\right\},\\
		2 R_j & =\min \left\{\left|X_j-X_k\right| :  k=1, \ldots, M \text { and } k \neq j\right\}. 
	\end{align*}
\end{lemma}

\begin{proof}  
%Let us consider   the Voronoi cells 
%$$\Gamma_j=\left\{x \in \mathbb{R}^d|| x-X_j|\leq| x-X_k \mid\right. \text{ for all }k \neq j, k=1, . ., M\}$$
%and the balls $B_j=\left\{ x \in \R^d | \left|x-X_j\right| \leq R_j\right\} \subset \Gamma_j$.   
For $0 < s < \frac{d}{2}$, by the one-body Hardy uncertainty estimate \eqref{one-body-hardy-unvertainty}, we have  
	\begin{equation*}
		\langle u, (-\Delta)^s_{|B} u \rangle  \ge  \mathcal{C}_{d,s}\int_{B} \frac{|u(x)|^2}{|x|^{2s}} dx - C(d,s) \int_{B} |u(x)|^2dx,
	\end{equation*}
	where $B$ is the unit ball in $\R^d$. By translations and dilations, we obtain 
	\begin{equation}\label{kinetic-in-balls}
		\langle u, (-\Delta)^s_{|B_j} u \rangle \ge \mathcal{C}_{d,s}\int_{B_j} \frac{|u(x)|^2}{|x - X_j|^{2s}} dx - \frac{C(d,s)}{R_j^{2s}} \int_{B_j} |u(x)|^2 dx
	\end{equation}
	where $B_j=\left\{ x \in \R^d :  \left|x-X_j\right| \leq R_j\right\}$ 	for all $j=1,2,...,M$. Since the balls  $\{B_j\}_{j=1}^M$ are disjoint, by the definition of $V$ and \eqref{kinetic-in-balls}, we have the operator inequality 
	$$
	(-\Delta)_{\R^d}^s -\beta V \ge (1 - \beta \mathcal{C}_{d,s}^{-1})(-\Delta)_{\R^d}^s -\lambda W
	$$
with 
	\begin{align*}
	W = \sum_{j=1}^M W_j,\quad 
		W_j(x) = \begin{cases}
			|x - X_j |^{-2s}  \quad & \text{ if }  |  x - X_j| > R_j,\\
			C(d,s) R_j^{-2s} \quad &\text{ if } |x - X_j| \le R_j.
\end{cases}
	\end{align*}
	By the (fractional) fermionic Lieb--Thirring inequality (see e.g. \cite{FLW-23}), for every density matrix $0 \le \gamma \le 1$,  we have 
	\begin{align*}
	\Tr \Big( ((-\Delta)_{\R^d}^s -\beta V)\gamma\Big) &\ge \Tr \Big(  ((1 - \beta \mathcal{C}_{d,s}^{-1})(-\Delta)_{\R^d}^s -\lambda W)\gamma\Big)\\
	&\ge - C_1(d,s) \beta^{1 + \frac{d}{2s}}  \int_{\R^d} W^{1 + \frac{d}{2s}}(x) dx \ge - C(d,s)  \beta^{1 + \frac{d}{2s}}  \sum_{j=1}^M R_j^{-2s}. 
	\end{align*}
%	
%	$$ {\rm Tr}(\gamma|p|^{2s}) = (1 -  \beta \mathcal{C}_{d,s}^{-1}) {\rm Tr}(\gamma|p|^{2s})  + \beta \mathcal{C}_{d,s}^{-1}  {\rm Tr}(\gamma|p|^{2s}).$$
%	, we have
%	\begin{align}
%		{\rm Tr }(\gamma (|p|^{2s} - V)) \ge {\rm Tr} \gamma((1 - \beta \mathcal{C}_{d,s}^{-1})|p|^{2s} -\beta W ),\label{tr-1-lambda-gamma-W}
%	\end{align}
%	
%	Let us recall  the fractional Lieb-Thirring inequality 
%	\begin{equation}
%		{\rm Tr} \gamma( K |p|^{2s} - \beta \mathcal{V} )  \ge - C(d,s) K^{-\frac{d}{2s}} \beta^{1 + \frac{d}{2s}} \int \mathcal{V}^{1 + \frac{d}{2s}} dx, \text{ for } 0 \le \mathcal{V}  \in L^{1 + \frac{d}{2s}}(\R^d),
%	\end{equation} 
%	Therefore, the inequality \eqref{tr-1-lambda-gamma-W} yields (note that $\beta$ will be small)
%	\begin{equation}\label{LT-for-gamma-W}
%		{\rm Tr} \gamma((1 - \beta \mathcal{C}_{d,s}^{-1})|p|^{2s} -\beta W ) \ge -C(d,s) \beta^{1 + \frac{d}{2s}}  \int_{\R^d} W^{1 + \frac{d}{2s}}(x) dx.
%	\end{equation}
%	By  a simple computation, we find that
%	$$ \int_{\R^d} W^{1 + \frac{d}{2s}} (x) dx \le  C_1(d,s) \sum_{j=1}^M R_j^{-2s}. $$
%	Thus, \eqref{inequality-Tr-gamma-ge-alpha-R-j}   follows  by  \eqref{LT-for-gamma-W}. Finally, we finish the proof of the Lemma. 
\end{proof}
Now,  we can provide 
\begin{proof}[Proof of  \eqref{eq:theo-3}] Since the cases $N=1$ and $N=2$ are easy, let us consider the case $N \ge 3$.  We write $N=M+L$ with $1 \le  L \le  N-2$ and $M =N- L$.  Let $P = (\pi_1,\pi_2)$  be any partition of $\{1,...,N\}$ with $|\pi_1| =L$ and $|\pi_2|=M$.   For each  $ i  \in \{1,...,N\}$, we define
%	\begin{equation}\label{defi-delta-i-pi-2}
%		\delta_i(\pi_2)  = \min\{  |x_i - x_j| \text{ for } j \in \pi_2, \text{ and } j \ne i \text{ if } i \in \pi_2  \}.
%	\end{equation}
	\begin{equation}\label{defi-delta-i-pi-2}
		\delta_i(\pi_2)  = \min\{  |x_i - x_j| \text{ for } j \in \pi_2 \backslash\{i\} \}.
	\end{equation}
	By Lemma \ref{lemma-Tr-gamma-Delta-s-V}, the $L$-body operator
	$$ h_P := \sum_{i \in \pi_1} (-\Delta_i)^{s} - \beta \sum_{i \in \pi_1} \delta_i^{-2s}(\pi_2)   + \alpha \sum_{i \in \pi_2} \delta_i^{-2s}(\pi_2) $$
	is non-negative on the anti-symmetric subspace of $L^2(\R^{dL})$ if 
	$$0<\beta<\mathcal{C}_{d,s},\quad \alpha\ge C(d,s)\beta^{1+\frac d {2s}}$$
	with a large universal constant $C(d,s)$. Next, we define the $N$-particle operators 
	\begin{align}
		H & = \binom{N}{L}^{-1} \frac{N}{L} \sum_{P} h_P, \quad 		\hat H  = \sum_{i=1}^N (-\Delta_i)^{s} - C_1 \sum_{i=1}^N \delta_i^{-2s}. 
	\end{align}
	By the definition of $\delta_i$, it is obvious that 
	$ \delta_i^{-2s}(\pi_2)  \le \delta_i^{-2s},   $ and hence
	\begin{align*}
		&\binom{N}{L}^{-1} \frac{N}{L} \sum_{P} 	\sum_{i  \in \pi_2 } \alpha \delta_i^{-2s}(\pi_2) \le  \binom{N}{L}^{-1} \frac{N}{L}  \binom{N-1}{L}   \sum_{i=1}^N \alpha \delta_i^{-2s} = \alpha \frac{(N-L)}{L} ,\\
		&\binom{N}{L}^{-1} \frac{N}{L} \sum_{P} 	\sum_{i  \in \pi_1 }\beta \delta_i^{-2s}(\pi_2)
		 \le 
		\binom{N}{L}^{-1} \frac{N}{L}  \binom{N-2}{L-1} \sum_{i=1}^N \beta \delta_i^{-2s}  =  \beta \frac{(N-L)}{N-1} \sum_{i=1}^N \delta_i^{-2s}.
	\end{align*}
	Therefore, we have the desired operator inequality 
	$\hat H \ge H \ge 0$ provided that  
	\begin{equation}\label{condi-N-L}
	 (N-L) \left(  \beta (N-1)^{-1}  -\alpha L^{-1} \right) \ge C_1.
	\end{equation}
	We can choose $L=[N/2]$, $\alpha=C(d,s)\beta^{1+\frac d {2s}}$ and $\beta=\beta(d,s) \in (0,\mathcal{C}_{d,s})$ sufficiently small, so that  \eqref{condi-N-L} holds for a small universal constant $C_1=C_1(d,s)>0$. The proof of   \eqref{eq:theo-3} is complete. 
	\end{proof}


\begin{thebibliography}{10}
	
	
	\bibitem{Adams75}
	{\sc R. A. Adams}.  {\em Sobolev Spaces}. Academic Press, Boston, 1975.
	
	
	\bibitem{AFAP03}
	{\sc R. A. Adams and J. J. Fournier}. {\em Sobolev spaces}, Academic Press, Vol. 140
	2003.
	
	\bibitem{Besicovitch-45} {\sc A. S. Besicovitch}. A general form of the covering principle and relative differentiation
of additive functions. {\em Proc. Cambridge Philos. Soc.} 41 (1945), pp. 103--110.


	
	\bibitem{Dyson-57} {\sc F. J. Dyson}. Ground-state energy of a hard-sphere gas. {\em Phys. Rev.} 106 (1957), pp.
20--26.
	
	\bibitem{DyLe-67} {\sc F.~J. Dyson and A.~Lenard}.  Stability of matter. {I}. {\em J. Math. Phys.}  8 (1967), pp.~423--434.
	
	
	\bibitem{EkhFra-06}
	{\sc T.~Ekholm and R.~L.~Frank}.
	{On Lieb-Thirring inequalities for Schr\"odinger operators with virtual level}.
	{\em Commun. Math. Phys.} 264 (2006), no. 3, pp.~725--740. 
	

\bibitem{FefdeL-86} {\sc C. Fefferman and R. de la Llave.} Relativistic stability of matter. {\em Revista Matem\'atica Iberoamericana} 2 (1986), pp. 119--213.
	
	
	\bibitem{Frank-09}
	{\sc R.~L.~Frank}.
	{A simple proof of Hardy-Lieb-Thirring inequalities},
	{\em  Commun. Math. Phys.} 290 (2009), pp.~789--800.
	
	\bibitem{FraLieSei-08}
	{\sc R.~L.~Frank, E.~H.~Lieb and R.~Seiringer}.
	 Hardy-Lieb-Thirring inequalities for fractional Schr{\"o}dinger operators.
	{\em J. Amer. Math. Soc.} 21 (2008), pp.~925--950.


	\bibitem{FrSe-12}
	{\sc R.~L. Frank and R.~Seiringer}. {Lieb--Thirring inequality for a model
		  of particles with point interactions}. {\em J. Math. Phys.} 53 (2012), pp.~095201.
	
	
	\bibitem{FLW-23}  {\sc R. L. Frank, A. Laptev and T. Weidl.} {\em Schr\"odinger operators: Eigenvalues and Lieb-Thirring
	Inequalities}. Cambridge University Press, 2023. 
	
		
	\bibitem{KogNam-21} {\sc K. K\"ogler and P.T. Nam.} The Lieb-Thirring inequality for interacting systems in strong-coupling limit. {\em Arch. Ration. Mech. Anal.} 240 (2021), pp. 1169--1202. 
	
	
	
	\bibitem{LaLu-18} {\sc S.~{Larson} and D.~{Lundholm}}. {Exclusion bounds for extended anyons}. {\em Arch. Rational Mech. Anal.} 227 (2018), pp. 309--365.
	%
	%
	\bibitem{LaLuNa-19} {\sc S.~{Larson}, D.~{Lundholm}, and P.~T. {Nam}}.  Lieb-Thirring  inequalities for wave functions vanishing on the diagonal set. {\em Annales Henri Lebesgue} 4 (2021), pp. 251--282. 


\bibitem{DyLe-68} {\sc A.~Lenard and F.~J. Dyson}.  Stability of matter. {II}, {\em J. Math. Phys.} 9 (1968), pp.~698--711.

	
	\bibitem{LiLo-01} {\sc E. H. Lieb and M. Loss}. {\em Analysis}. Second edition, Graduate Studies in Mathematics,
	American Mathematical Society, Providence, RI, 2001.
	
	\bibitem{LiSe-10} {\sc E.~H. Lieb and R.~Seiringer}. {\em The stability of matter in quantum mechanics}. Cambridge University Press, 2010.
	
	\bibitem{LiTh-75}
	{\sc E.~H. Lieb and W.~E. Thirring}. {{Bound for the Kinetic Energy of
			Fermions Which Proves the Stability of Matter}}. {\em Phys.\ Rev.\ Lett.} 35
	(1975), pp.~687--689.
	
	\bibitem{LiTh-76} {\sc E.~H. Lieb and W.~E. Thirring.} Inequalities for the moments of the eigenvalues of the
Schr\"odinger Hamiltonian and their relation to Sobolev inequalities. In: {\em Studies in Mathematical
Physics}, Princeton University Press, NJ, USA, 1976, pp. 269--303.
	
	\bibitem{LieYau-88}   {\sc E. H. Lieb and H.-T. Yau.}   The stability and instability of relativistic matter. {\em Commun. Math. Phys.} 118 (1988), pp. 177--213. 

	

		
	
		
	\bibitem{LuNaPo-16}
	{\sc D.~Lundholm, P.~T. Nam, and F.~Portmann}, {Fractional
		{H}ardy-{L}ieb-{T}hirring and related inequalities for interacting systems},
	{\em Arch. Ration. Mech. Anal.} 219 (2016), pp.~1343--1382.
	
	\bibitem{LuPoSo-15}
	{\sc D.~Lundholm, F.~Portmann, and J.~P. Solovej}, {Lieb--{T}hirring bounds
		for interacting {B}ose gases}. {\em Commun. Math. Phys.} 335 (2015), pp.~1019--1056.
	
	
	
	\bibitem{LuSe-18} {\sc D. Lundholm and R. Seiringer}. Fermionic behavior of ideal anyons. {\em Lett. Math. Phys.} 108 (2018), pp.  2523-2541. 
	
	
	\bibitem{LuSo-13}
	{\sc D.~Lundholm and J.~P. Solovej}. {Hardy and {L}ieb--{T}hirring
		inequalities for anyons}. {\em Commun. Math. Phys.} 322 (2013), pp.~883--908.

	
	
	\bibitem{LuSo-13b} {\sc D. Lundholm and J.P. Solovej.}  {\em Local exclusion principle for identical particles obeying intermediate and fractional statistics}. {\em 	 Phys. Rev. A} 88 (2013), 062106.
	
	\bibitem{LuSo-14} {\sc D. Lundholm and J.P. Solovej}. {\em Local exclusion and  {L}ieb-{T}hirring inequalities for intermediate and fractional statistics}. 
	  {\em Ann. Henri Poincar\'{e}} 15 (2014), pp.~1061--1107.
	 

	
	
	
	\bibitem{Morse-47}   {\sc A. O. Morse.} Perfect blankets. {\em Trans. Am. Math. Soc.} 61 (1947),  pp. 418--442.
	
	
	\bibitem{Nam-17} {\sc P.T. Nam.} Lieb-Thirring inequality with semiclassical constant and gradient error term. {\em J. Funct. Anal.} 274 (2018), pp. 1739--1746.
	
	\bibitem{Nam-22} {\sc P.T. Nam.}  A proof of the Lieb-Thirring inequality via the Besicovitch covering lemma. {\em Acta Mathematica Vietnamica} 48 (2023), pp. 75--81.
	
		
	\bibitem{Leoni-book2017t}
	{\sc G. Leoni}. \emph{A First Course in Sobolev Spaces}, 2nd ed. American Mathematical Society, 2017.
	
	\bibitem{leoni2023}
	{\sc G. Leoni}.   \emph{A First Course in Fractional Sobolev Spaces}. American Mathematical Society, 2023.
	
	\bibitem{Herbst-CMP-1997}
	{\sc I. W. Herbst}.  Spectral theory of the operator    $   (p^2 +m^2)^\frac{1}{2} -Ze^2/r$. {\em Commun. Math. Phys.} 53 (1977), pp. 285--294. 
	
%	\bibitem{R76} {\sc G. V. Rozenblum}. Distribution of the discrete spectrum of singular differential operators. {\em Izvestiya Vysshikh Uchebnykh Zavedenii Matematika} 1 (1976), pp. 75--86 .
	
	
%	\bibitem{LSSY05}
%	{\sc E.~H. Lieb, R. Seiringer, J. P. Solovej, and J. Yngvason, \emph{The Mathematics of the Bose Gas and its Condensation}, \relax Springer,  2005.
%	%\bibitem{Nam-18} {\sc P.T. Nam}, {\em Lieb--Thirring inequality with semiclassical constant and gradient error term}, J. Funct. Anal. 274 (2018), pp. 1739--1746.
	%
	%\bibitem{SolSorSpi-10}
	%	{\sc J.~P.~Solovej, T.~\O .~S\o rensen and W.~L.~Spitzer},
	%	{\em Relativistic Scott correction for atoms and molecules},
	%	Comm. Pure and Applied Math., 63 (2010), pp.~39--118.
	
	%\bibitem{Tr-78}
	%{\sc H.~Triebel}, {\em Interpolation theory, function spaces, differential
		%  operators}, vol.~18 of North-Holland Mathematical Library, North-Holland
	%  Publishing Co., Amsterdam-New York, 1978.
	
	
\end{thebibliography}
\end{document}